\documentclass[structabstract]{aa}
\usepackage{amsmath,amsfonts,amssymb}
\usepackage{xspace}
\usepackage[dvipsnames,usenames]{color}
\usepackage{graphicx}
\usepackage{txfonts}
\usepackage{color}

\makeatletter
\let\NAT@parse\undefined
\makeatother

\usepackage{natbib}

\usepackage{multirow}
\bibpunct{(}{)}{;}{a}{}{,} 

\definecolor{LinkColor}{rgb}{0.0,0.2,0.5}
\usepackage[pdfpagemode=UseNone,pdfstartview=FitB]{hyperref}
\hypersetup{
  breaklinks=true,
  colorlinks=true,
  linkcolor=LinkColor,
  citecolor=LinkColor,
  filecolor=LinkColor,
  menucolor=LinkColor,
  pagecolor=LinkColor,
  urlcolor=LinkColor,
}

\makeatletter
\newcommand{\MathFuncName}[1]{{\operator@font #1}}
\newcommand{\MathFunc}[1]{\mathop{\operator@font #1}\nolimits}
\newcommand{\MathFuncWithLimits}[1]{\mathop{\operator@font #1}\limits}
\makeatother

\newcommand{\Tag}[1]{\mathsf{#1}}        
\newcommand{\V}[1]{\boldsymbol{#1}}      
\newcommand{\M}[1]{\mathbf{#1}}          

\newcommand{\mathe}{\mathrm{e}}
\newcommand{\mathi}{\mathrm{i}}

\newcommand{\Reals}{\mathbb{R}}

\newcommand{\ie}{\emph{i.e.}\xspace}

\newcommand{\etc}{\emph{etc.}\xspace}

\makeatletter

\newcommand{\@SymbolBuilderWithLabel}[4]{%
  \begingroup \escapechar\m@ne\xdef\my@tempa{\string#1}\endgroup
  \expandafter\@ifundefined{\my@tempa}{%
    \def\my@tempb{[} 
    \expandafter\edef\csname\my@tempa\endcsname{%
      \noexpand\@ifnextchar\my@tempb%
          {\csname\my@tempa @b\endcsname}%
          {\csname\my@tempa @a\endcsname}%
    }
    \expandafter\def\csname\my@tempa @a\endcsname{%
      {#2}_{#4}%
    }
    \expandafter\def\csname\my@tempa @b\endcsname[##1]{%
      {{#3}^{#4}_{##1}}%
    }
  }{\@latex@error{\noexpand#1is already defined}\@ehc}%
}
\newcommand{\@SymbolBuilder}[3]{
  \begingroup \escapechar\m@ne\xdef\my@tempa{\string#1}\endgroup
  \expandafter\@ifundefined{\my@tempa}{%
    \def\my@tempb{[}
    \expandafter\edef\csname\my@tempa\endcsname{%
      \noexpand\@ifnextchar\my@tempb%
          {\csname\my@tempa @b\endcsname}%
          {\csname\my@tempa @a\endcsname}%
    }
    \expandafter\def\csname\my@tempa @a\endcsname{%
      #2%
    }
    \expandafter\def\csname\my@tempa @b\endcsname[##1]{%
      {{#3}_{##1}}%
    }
  }{\@latex@error{\noexpand#1is already defined}\@ehc}%
}

\newcommand{\SymbolWithLabel}[3]{%
  \@SymbolBuilderWithLabel{#1}{#2}{#2}{#3}}
\newcommand{\Symbol}[2]{%
  \@SymbolBuilder{#1}{#2}{#2}}

\newcommand{\VectorSymbolWithLabel}[3]{%
  \@SymbolBuilderWithLabel{#1}{\V{#2}}{#2}{#3}}
\newcommand{\VectorSymbol}[2]{%
  \@SymbolBuilder{#1}{\V{#2}}{#2}}

\newcommand{\MatrixSymbolWithLabel}[3]{%
  \@SymbolBuilderWithLabel{#1}{\M{#2}}{#2}{#3}}
\newcommand{\MatrixSymbol}[2]{%
  \@SymbolBuilder{#1}{\M{#2}}{#2}}

\makeatother

\newcommand{\uv}{$(u,v)$\xspace}

\newcommand{\Param}{x}
\newcommand{\VParam}{\V{\Param}}

\newcommand{\Grid}[1]{\stackrel{%
    {\fboxsep0.09em{\vbox{\hbox{\fbox{}}\vskip0pt}}}}{#1}}

\newcommand{\OTF}{G} 

\Symbol{\Wavelength}{\lambda}
\Symbol{\GridWavelength}{\Grid{\Wavelength}}

\newcommand{\DirLetter}{a}
\Symbol{\Dir}{\V{\DirLetter}}
\Symbol{\GridDir}{\Grid{\V{\DirLetter}}}

\newcommand{\ParamLetter}{x}
\VectorSymbol{\X}{\ParamLetter}

\MatrixSymbol{\XForm}{T}
\MatrixSymbol{\OTFop}{\OTF}
\MatrixSymbol{\DFTop}{F}
\MatrixSymbol{\InterpOp}{R}
\MatrixSymbol{\ModelOp}{A}
\MatrixSymbol{\PSFop}{H}
\MatrixSymbol{\ApodizationOp}{S}

\Symbol{\PSF}{h}
\VectorSymbolWithLabel{\Xpsf}{x}{\Tag{psf}}

\newcommand{\DataLetter}{y}
\VectorSymbol{\Data}{\DataLetter}

\newcommand{\HyperLetter}{\theta}
\VectorSymbol{\Hyper}{\HyperLetter}

\VectorSymbol{\Model}{m}

\VectorSymbol{\Error}{e}

\VectorSymbol{\Residual}{r}

\newcommand{\DataTag}{\Tag{data}}
\newcommand{\PriorTag}{\Tag{prior}}

\newcommand{\ErrorTag}{\Tag{err}}

\MatrixSymbolWithLabel{\Cerror}{C}{\ErrorTag}
\MatrixSymbolWithLabel{\Werror}{W}{\ErrorTag}
\MatrixSymbolWithLabel{\Cprior}{C}{\PriorTag}
\MatrixSymbolWithLabel{\Wprior}{W}{\PriorTag}
\MatrixSymbol{\Identity}{I}

\newcommand{\Fcost}{f}

\newcommand{\Fdata}{\Fcost_\DataTag}
\newcommand{\Fprior}{\Fcost_\PriorTag}

\newcommand{\Mira}{MiRA\xspace}

\newcommand{\mwc}{HD$\,$163296}

\definecolor{FireBrick}{rgb}{0.70,0.13,0.13}

\definecolor{SeaGreen}{rgb}{0.13,0.70,0.13}

\begin{document}
\title{Milli-arcsecond images of the Herbig Ae star
  \mwc}

\author{S.\ Renard\inst{1} \and F.\ Malbet\inst{1} \and M.\
  Benisty\inst{2} \and E.\ Thi\'ebaut\inst{3} \and J-P.\
  Berger\inst{4,1}} %
\institute{%
  Laboratoire d'Astrophysique de Grenoble, CNRS-UJF UMR5571, BP 53,
  F-38041 Grenoble, France --\\ \email{Stephanie.Renard,
    Fabien.Malbet@obs.ujf-grenoble.fr} \label{laog} %
  \and INAF-Osservatorio Astrofisico di Arcetri, Largo E. Fermi 5,
  50125 Firenze, Italy -- \email{benisty@arcetri.astro.it} \label{oaa}%
  \and Centre de Recherche Astrophysique de Lyon, CNRS-UCBL-ENSL
  UMR5574, F-69561 St-Genis-Laval, France --\\
  \email{thiebaut@obs.univ-lyon1.fr} \label{cral} %
  \and European Southern Observatory, Alonso de Cordova, 3107,
  Vitacura, Chile -- \email{jpberger@eso.org} \label{eso} %
}

\date{Received ; accepted }



\abstract
{The very close environments of young stars are
the hosts  of fundamental physical processes, such as planet
formation, star-disk interactions, mass accretion, and ejection. The complex
morphological structure of these environments has been confirmed
by the now quite rich data sets obtained for a few objects by near-infrared long-baseline interferometry.} %
{We gathered numerous interferometric measurements for the young
star \mwc\ with various interferometers (VLTI, IOTA, KeckI and
CHARA), allowing for the first time an image independent of any
a priori model to be reconstructed.} %
{Using the Multi-aperture image Reconstruction Algorithm (\Mira), we reconstruct images of \mwc\ in
the $H$ and $K$ bands. We compare these images with reconstructed images obtained from simulated data using a physical model of the environment of \mwc.} %
{We obtain model-independent $H$ and $K$-band images of the
surroundings of \mwc. The images present several significant
features that we can relate to an inclined
asymmetric flared disk around \mwc\ with the strongest intensity
at  about  4-5\,mas. Because of the  incomplete spatial  frequency
coverage, we cannot state whether each of them individually is peculiar in any way.} %
{For the first time, milli-arcsecond images of the environment of a
young star are produced. These images confirm that the morphology of
the close environment of young stars is more complex
than the simple models used in the literature so far.} %



\maketitle


\section{Introduction}
\label{sec:introduction}

The process of star formation triggered by the collapse and
fragmentation of a molecular cloud leads to the birth of a young star
surrounded by a circumstellar disk and outflows. The disks are believed to be the place where the planets form
\citep{Boss1997, Mayer2002}, and are composed of a mixture of
gas and dust with a wide range of grain composition
\citep{HenningMeeus2009}. The standard picture of the close
environment of pre-main sequence stars is so far limited to a
quasi-stationary accreting disk partially reprocessing the irradiation
from the central protostar with potential planetary gaps
opened by newly formed planets. Some of the accreted material is
ejected by means of bipolar outflows but their precise origin has not yet
been identified. Most of the models
are presently assumed to be symmetric around the star rotation axis and
stationary.  However  hydrodynamical  turbulence, gravitational
waves, magneto-rotational instabilities, vortices and
thermal instabilities occurring on the AU-scale are known to play
a major role  in star and planet  formation \citep[e.g.,][]{Balbus1991}.
The close environments of young stars are therefore
not expected to be as simple as they are currently modeled, but the
observational   measurements   do not provide tight enough constraints  to  unambiguously identify strong departures from axisymmetrical models.

The photometric and spectroscopic observations obtained with very modest spatial resolution are usually integrated over a subarcsecond field of view which corresponds to several tens of AUs  at the distance of the closest star formation regions (140pc). Observations at the scale of 1\,AU and below correspond to 7 milli-arcseconds (mas) angular scale and requires therefore optical long-baseline interferometry.
However, even if  long-baseline    optical
interferometry is capable of reaching very high angular resolution, the observations
are usually limited to a small number of measurements
\citep[see][for a review]{MillanGabet2007}. With the advent of interferometers with more
than two telescopes and with significantly high spectral resolution
(e.g.,   VLTI/AMBER,  CHARA/MIRC),   the  number   of  interferometric
observations  has  significantly  increased  allowing  the  first
images to be reconstructed with aperture synthesis techniques.  Images
of stellar surfaces \citep[e.g.,][]{Monnier2007, Hautbois2009, Zhao2009}, binaries
\citep[e.g.,][]{Zhao2008,Kraus2009},
or circumstellar shells around
evolved stars \citep[e.g.,][]{LeBouquin2009} have been obtained mostly
for objects brighter than the brightest young stars. In the young stellar object field, we are at a comparable stage
to that reached 40 years ago when the first 3 antennas of the VLA became operational
producing the first radio-interferometry images
\citep{Hogg1969}. We report the
first attempt  to reconstruct  an image of a  circumstellar disk
around a young star using mostly
AMBER/VLTI interferometric data.

In this study, we present reconstructed images of the young star \mwc\ (MWC\,275), an isolated Herbig Ae star (HAe) of
spectral type A1, with a $\sim$30\,L$_{\sun}$ luminosity, and a $\sim$2.3\,M$_{\sun}$
mass located at $122^{+17}_{-13}$\,pc \citep{vandenAncker1998,
  Natta2004, Montesinos2009}.
In scattered light \citep{Grady2000} and at millimeter wavelengths
\citep{Mannings1997}, a disk has been detected on large scales, traced out
to  540\,AU.  The CO millimeter  line  observations  have  revealed a
  large-scale inclined disk in Keplerian rotation probably
evolving towards a debris disk phase \citep{Isella2007}.
\mwc\ also exhibits an asymmetric outflow
perpendicular to the disk, with a chain of six Herbig-Haro knots
(HH409) tracing the history of mass loss \citep{Devine2000,
  Wassell2006}. The emission of the innermost regions, observed in far-UV emission
lines, have been attributed to optically thin gas accreting onto the
stellar surface, a magnetically confined wind, or shocks at the base
 of the jet \citep{Deleuil2005, Swartz2005}.
\mwc\ has been observed with several interferometers \citep[see][B10 hereafter]{Benisty2010}.
Using the largest set of interferometric data of a young star, we present here the
first reconstructed images of a complex young stellar object.

The article is organized as follow. Sect.~\ref{sec:method} describes
the image reconstruction method and the methodology employed to extract the best reconstructed image.
Sect.~\ref{sec:results+analysis} presents the reconstructed images
obtained in the $H$ and $K$ bands and their analysis using simulated
data    generated    from     inner    disk    models.    In
Sect.~\ref{sec:discussion}, we describe the  choices made during the image
reconstruction  process,  and  discuss  the physical  meaning  of  the
features seen in the image, as well as the consequences for the models
commonly used. Finally, Sect.~\ref{sec:conclusion}
summarizes our results and provides some perspectives for the future.


\section{Image reconstruction}
\label{sec:method}

\subsection{Image reconstruction by \Mira}
\label{sec:MiRA}

The principle of interferometry is to interfere coherently the light coming
from a single astronomical source from two or more independent telescopes
\citep{Lawson2000,Malbet2007}.  An interferometer measures a complex number
referred to as the \emph{visibility}. According to the Van Cittert-Zernicke
theorem, this complex visibility, $V\,\mathe^{\mathi\,\phi}$, is the Fourier
transform of the object brightness distribution at the spatial frequency of
the observations, given by the projected baseline in units of wavelength
($\mathrm{B}/\lambda$).  The visibility amplitude, $V$, is related to the
spatial extent of the emission, while the phase, $\phi$, provides the location
of the photocenter. However, the requested infrastructure to carry out optical
interferometry is complex and has led to a limitation in the number of
telescopes \citep{Baldwin2002}.  The main consequence is to provide a sparse
sampling of the spatial frequencies, the so-called \uv plane.  Moreover, the
absolute value of the phase $\phi$ is lost due to atmospheric turbulence that
randomly modifies it.  However, by adding the phases of the fringes measured
for each baseline over a 3-telescope configuration, one can measure an
additional quantity, the \emph{closure phase}, which is insensitive to the
atmospheric disturbance \citep{Monnier2003,Monnier2006}. The closure phase
includes part of the Fourier phase information, and is related to the global
asymmetry of the emission: a point-symmetric object has a zero closure phase.
The main observables are therefore the squared visibility amplitudes, $V^{2}$,
and the closure phases (CP).

The objective of the image reconstruction is to numerically retrieve an
approximation of the true brightness distribution of the source given the
interferometric measurements. To account for the data, the Fourier transform
of the sought image should fit the measured complex visibilities.  However,
due to the sparse \uv coverage, the image reconstruction problem is ill-posed
as there are more unknowns, e.g., the \emph{pixels} of the image, than
measurements.  Additional prior constraints are therefore required to
supplement the available data and retrieve a unique and stable solution.  A
very general solution is to define the optimal image to be the solution to the
optimization problem \citep{Thiebaut2005, Thiebaut2009}:
\begin{equation}
  \min_{\VParam} \left\{ \Fdata(\VParam) + \mu \, \Fprior(\VParam) \right\}
  \text{ with } \VParam \geqslant 0 \text{ and } \sum_n \VParam_n = 1
  \label{eq:img-prob}
\end{equation}
where $\VParam\in\Reals^N$ are the pixel values in the discretized image and $N$ the
number of pixels.  In problem (\ref{eq:img-prob}) the strict constraints
$\VParam \geqslant 0$ and $\sum_n \VParam_n = 1$ account for the
non-negativity and the normalization of the brightness distribution.  The
objective function in the optimization problem (\ref{eq:img-prob}) is a joint
criterion of two components:
\begin{itemize}
\item Minimization of the \emph{likelihood} term $\Fdata(\VParam)$, which enforces
  agreement of the sought image with the data.  In practice, this term is
  derived from the noise statistics. For instance, it is the $\chi^2$ of
  the data for Gaussian statistics.
\item Minimization of the \emph{regularization} term $\Fprior(\VParam)$, which favors
  images that are the \emph{simplest} or the \emph{smoothest} according to a priori
  assumptions.
\end{itemize}
The parameter $\mu>0$ is used to tune the relative importance of the two terms so as to select the most regular image among all those compatible with the data.

Many different algorithms have been developed to solve the image reconstruction problem
from optical interferometry data \citep[e.g.,][]{Cotton2008, Thiebaut2009}.
In this paper, we use the  Multi-Aperture Image Reconstruction Algorithm
\citep[\Mira; by][]{Thiebaut2008}.
\Mira is capable of dealing with  any available interferometric data (complex visibilities,
squared visibilities, closure phases, \etc), and has been successfully
used to process real data
\citep[e.g.,][]{Lacour2008, Lacour_et_al-2009-Chi_Cygni, LeBouquin2009, Hautbois2009}.
When phases are missed because of the  atmospheric turbulence, the \Mira algorithm
can directly fit the available interferometric observables without explicitly rebuilding
the missing phases. Finally, one can choose the most effective of various regularization methods integrated into \Mira for the type of object observed and check
the effect of the regularization on the resulting image. \Mira directly attempts
to solve the problem in Eq.~(\ref{eq:img-prob}) using an iterative  non-linear optimization
algorithm \citep{Thiebaut2002}. Because of the missing phases, the function to be  minimized
is not however convex thus has several local minima. The final image is therefore determined
by the data, the choice of the regularization (and its level), and the initial image.

\subsection{Methodology used in this work}
\label{sec:methodology}

Systematic tests have been performed on the \Mira algorithm by Renard
et al. (in prep. ; RTM10 hereafter) in which images of ten astrophysical
objects were reconstructed, for different \uv coverages and
signal-to-noise ratios. Twelve regularizations were tested and
images  were  reconstructed for  a  set of  weight factors
$\mu$. The optimal solution is the one that
minimizes the mean-squared distance between the model and the
reconstructed image. The RTM10 tests led to the following conclusions:
\begin{itemize}
\item The \emph{total variation} regularization \citep{Strong_Chan-2003-total_variation}, which minimizes the total norm of
  the image gradient, is the best regularization method in most of the cases.
\item The weight factor $\mu$ depends, within one order of
  magnitude, on neither the amount of data, the signal-to-noise ratio, nor
 the object type, but only on the type of the regularization
  used. Each regularization has its best value for $\mu$.
\end{itemize}
The chosen regularization for this work is therefore the \emph{total variation} and the weight factor $\mu$ is set to be $10^2$. The effect of
changing $\mu$ is discussed in Appendix~\ref{app:mu}.  As explained above,
because part of the phase information is lost, the image reconstruction from
squared visibilities and closure phases yields to a criterion with multiple
local minima. The solution depends on the initial image used to start the
iterative process. Different starting images, such as a Dirac or a random flux
distribution among pixels, were tested and, thanks to a good
regularization term and a sufficient number of data points, the solutions obtained were always found to be the same.  Here, we choose a symmetric Gaussian as a starting image.

Once the parameters of the image reconstruction process are set, the images
can be reconstructed. The solutions are not straightforward to analyze because of artifacts caused by the image reconstruction process or the
quality of the data set, e.g., voids in the \uv plane, error bars. There are
no objective criteria to distinguish between the actual structures from the
object and the artifacts caused only by the data structure. We therefore
performed a comparative analysis between our results and the results obtained
from simulated data from the B10 model for \mwc. To do so, we simulated fake
data sets, using the B10 model and the same \uv plane and errors as
in the real data set. Image reconstruction was also performed for the simulated
data in the same conditions and compared to the image model. This comparative
method is important to understand what could be trusted in the actual
reconstructed images and what could not.

\subsection{Interferometry data set}
\label{sec:dataset}

The principal characteristics of the data set is summarized in this
section. A more detailed description is given in the Appendix~\ref{app:data}. \mwc\ was observed with several
interferometers. The large data set comprises $H$ and $K$-band data
from VLTI \citep{Benisty2010}, IOTA \citep{Monnier2006}, Keck-I
\citep{Monnier2005}, and CHARA \citep{Tannirkulam2008}.  This data
set represents the largest set of interferometric data available so
far for a young stellar object. Since a trade-off has to be found
between enough data to pave the \uv plane and the wavelength
dependency of the observed object, we decided to reconstruct two
different images, one in $H$ and one in $K$ band, using all the
spectral channels available in each band \ie assuming the object to be grey in
each band (see Sect.~\ref{sec:wavelength} for further discussion).


\section{Results and first image analysis}
\label{sec:results+analysis}

\subsection{Reconstructed images in the $H$ and $K$ bands}
\label{sec:results}

\begin{figure*}
  \centering
  \hfill
  \includegraphics[height=6.7cm]{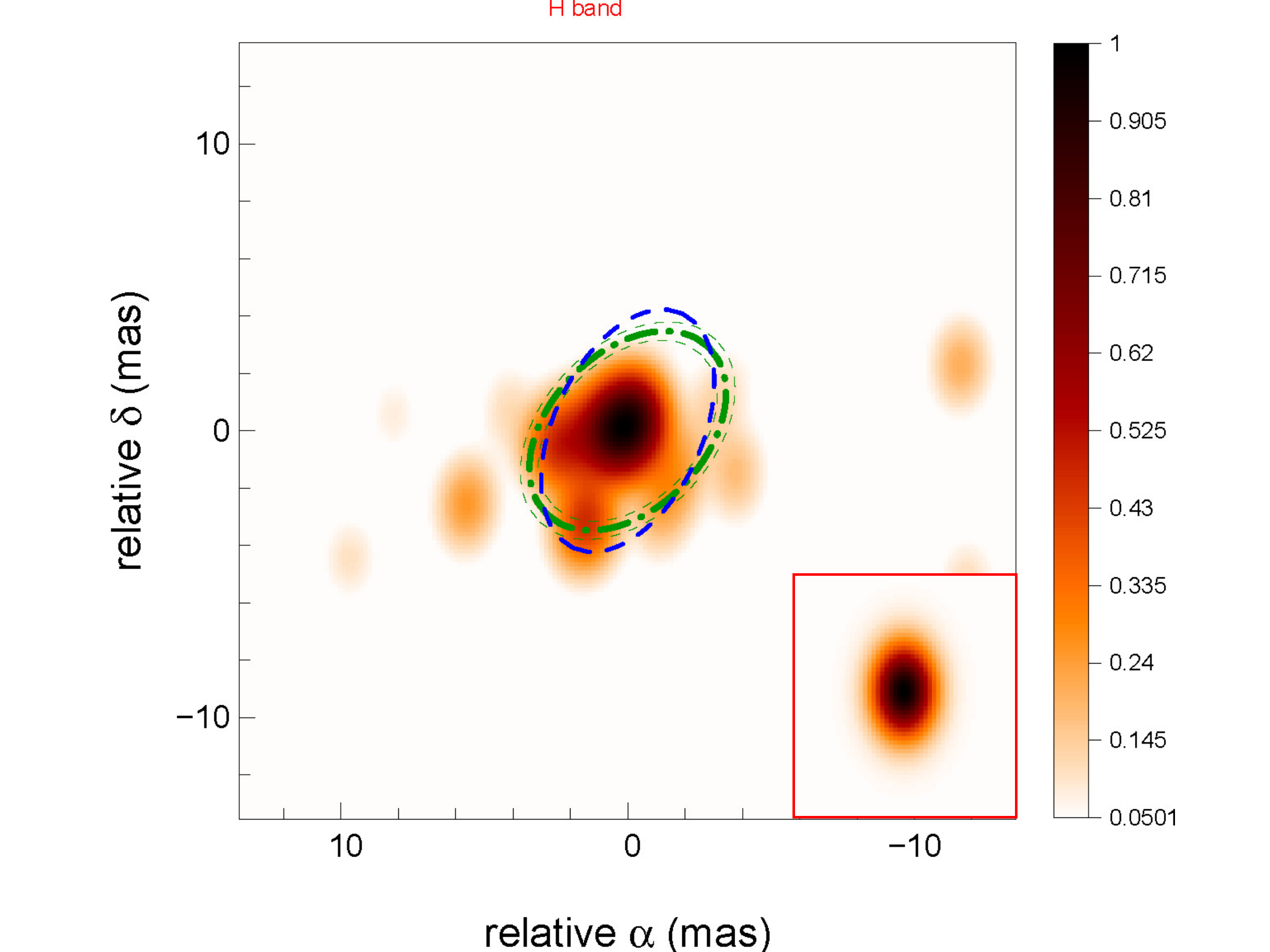}
  \hfill
  \includegraphics[height=6.7cm]{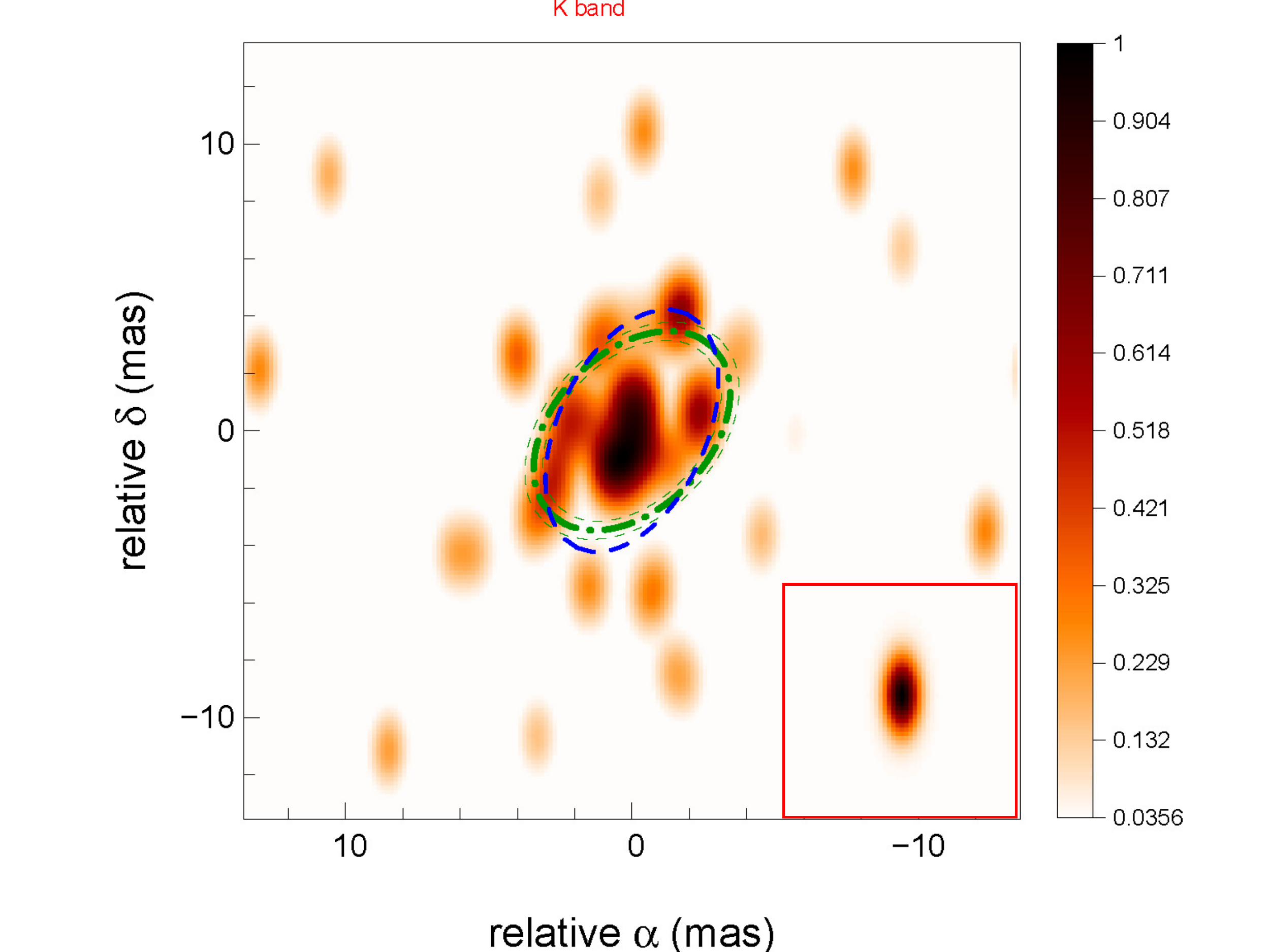}
 \hfill ~
 \caption{Reconstructed images of \mwc\ in the $H$ (left) and $K$ bands
   (right), after a convolution with a Gaussian beam at the
   interferometer resolution. The colors are scaled to the squared
   root of the intensity with a cut corresponding to the maximum
   expected dynamic range (see text for details). The blue ellipse
   traces the location of the main secondary blobs, and the green
   dot-dashed ellipse corresponds to the location of the rim in the B10
   model,   with   its   width  given   by   the     green dashed
   ellipses. North is up and east is left.  The sub-panel in the
   right corner of each  plot indicates the Gaussian beam at the
   interferometer            resolution,           applicable to
   Figs.~\ref{fig:model-img-rec},~\ref{fig:modelSansInt-img-rec},    and
   \ref{fig:pionier}.}
\label{fig:real-img-rec}
\end{figure*}
\begin{figure*}
  \centering
  \hfill
  \includegraphics[height=6.7cm]{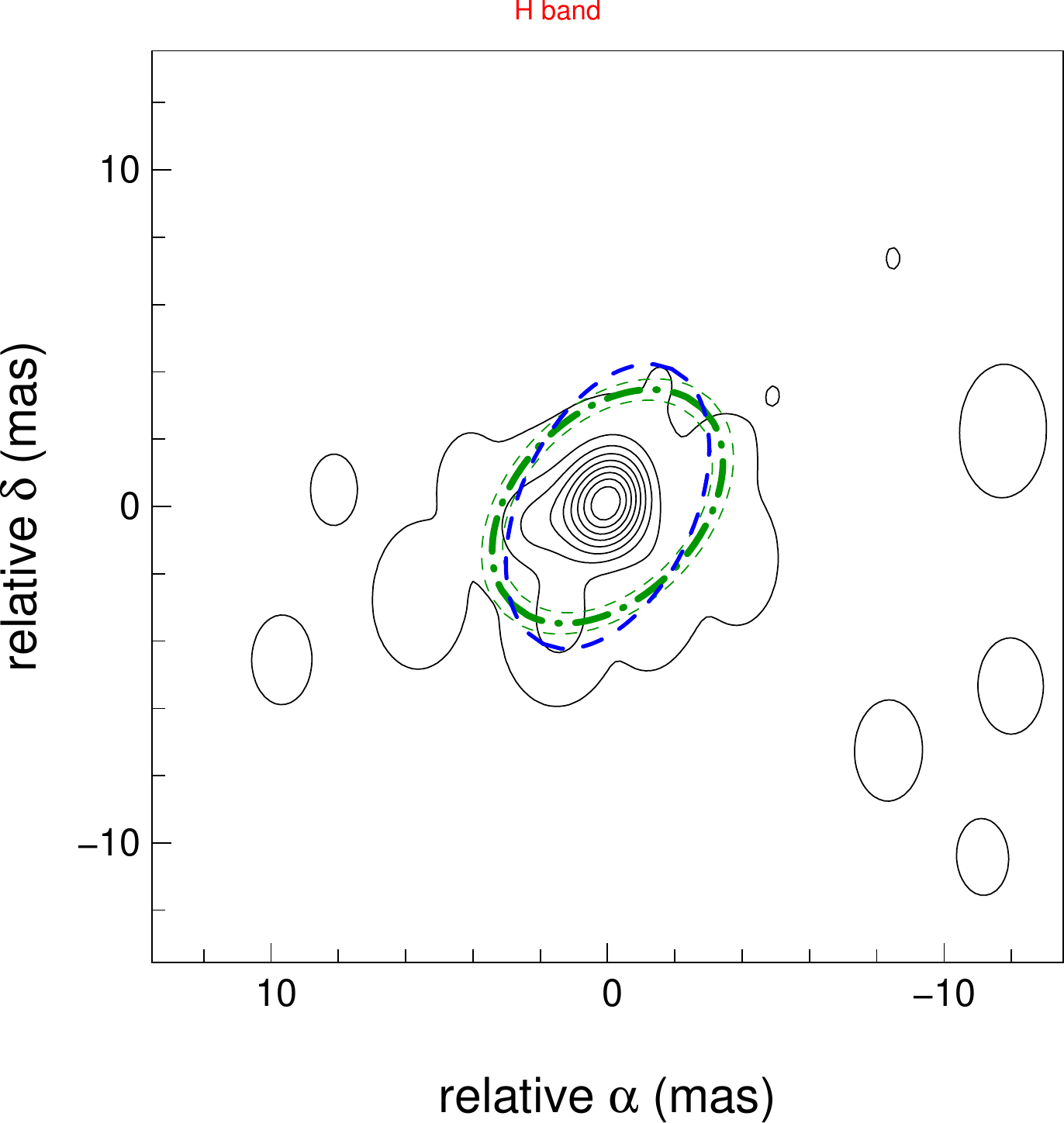}
  \hfill \hspace*{1.2cm}
  \includegraphics[height=6.7cm]{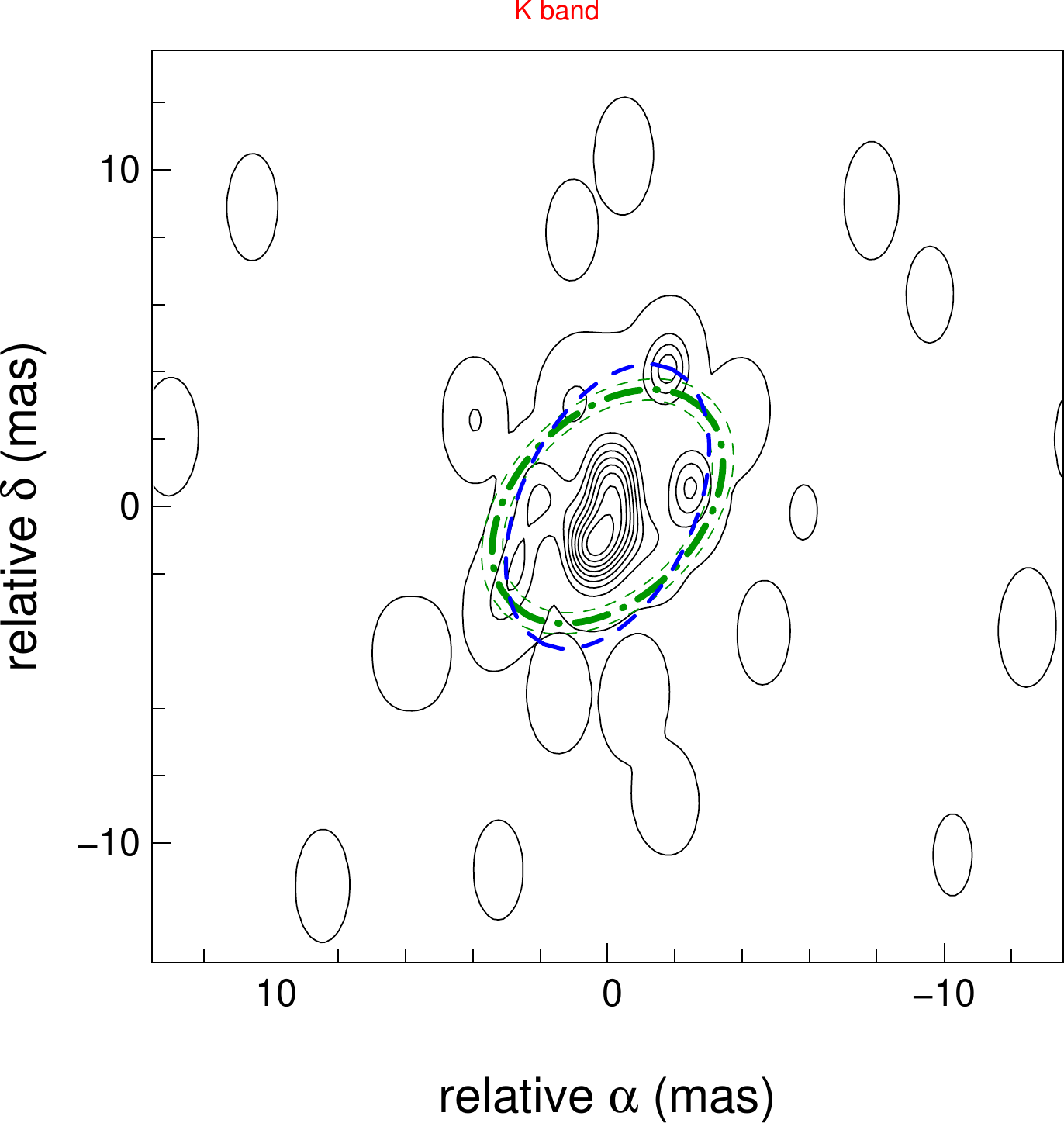}
  \hfill \hspace*{1.2cm} ~
  \caption{Contours of the reconstructed images of \mwc\ in the $H$
    (left) and $K$ bands (right), after a convolution with a Gaussian
    beam at the interferometer resolution. The contours vary linearly
    between the minimum cut corresponding to the maximum expected dynamic
    range and the image maximum, with a step around 0.1.}
  \label{fig:contour-img-rec}
\end{figure*}

Using the methodology described in Sect.~\ref{sec:methodology}, we
reconstruct the images of \mwc\ in the $H$ and $K$ bands. To
produce a \emph{rendu} similar to that usually used in
radio-interferometry, we convolve all the resulting images with a
Gaussian beam at the interferometer resolution defined by the \uv
plane.  The resulting images are plotted in
Fig.~\ref{fig:real-img-rec} with a color scale and in
Fig.~\ref{fig:contour-img-rec} with linear contours. As the
minimum cut, we use the level corresponding to the expected dynamic
range (see Sect.~\ref{sec:dynamics} for the discussion on how to
compute this value).

At first look, the spot representing the star is unambiguous and
corresponds to the maximum of the images in the two bands. Around this
central spot, one can see many secondary blobs. In the next section, we will discuss the level of confidence in these blobs by comparing
them to the results obtained on simulated data from the B10 model. The
main secondary blobs are concentrated in the center of the image
around the brightest spot at a distance of smaller than 4-5\,mas. To show  the  location  of these  secondary blobs, an ellipse is drawn using a dashed blue line in Figs.~\ref{fig:real-img-rec} and
\ref{fig:contour-img-rec} with a semi-major axis of 4.5\,mas, an
inclination   of   $\sim$55$\,\degr$,   and   a   position   angle   of
$\sim$155$\,\degr$. When considering their repartition along
the ellipse, we find that they  are less numerous at the bottom of the
ellipse. The percentage of flux in these blobs is around
30\% in the $K$ band and 24\% in the $H$ band.

Inside the ellipse, the central spot is not point-like. In the $H$
band, the central spot is extended, while in the $K$ band the energy
is spread into two close separated spots. The emission does not decrease slowly
from the central spot towards the exterior but instead shows a rapid decay before
increasing again when crossing the ellipse to finally decrease at large
distances. We also note that the ellipse and the central spots are not
exactly centered, although this result may not be relevant.

\subsection{Comparison with simulated images from models}
\label{sec:analysis}

\begin{figure*}[t]
  \centering
  \hfill
  \includegraphics[height=6.7cm]{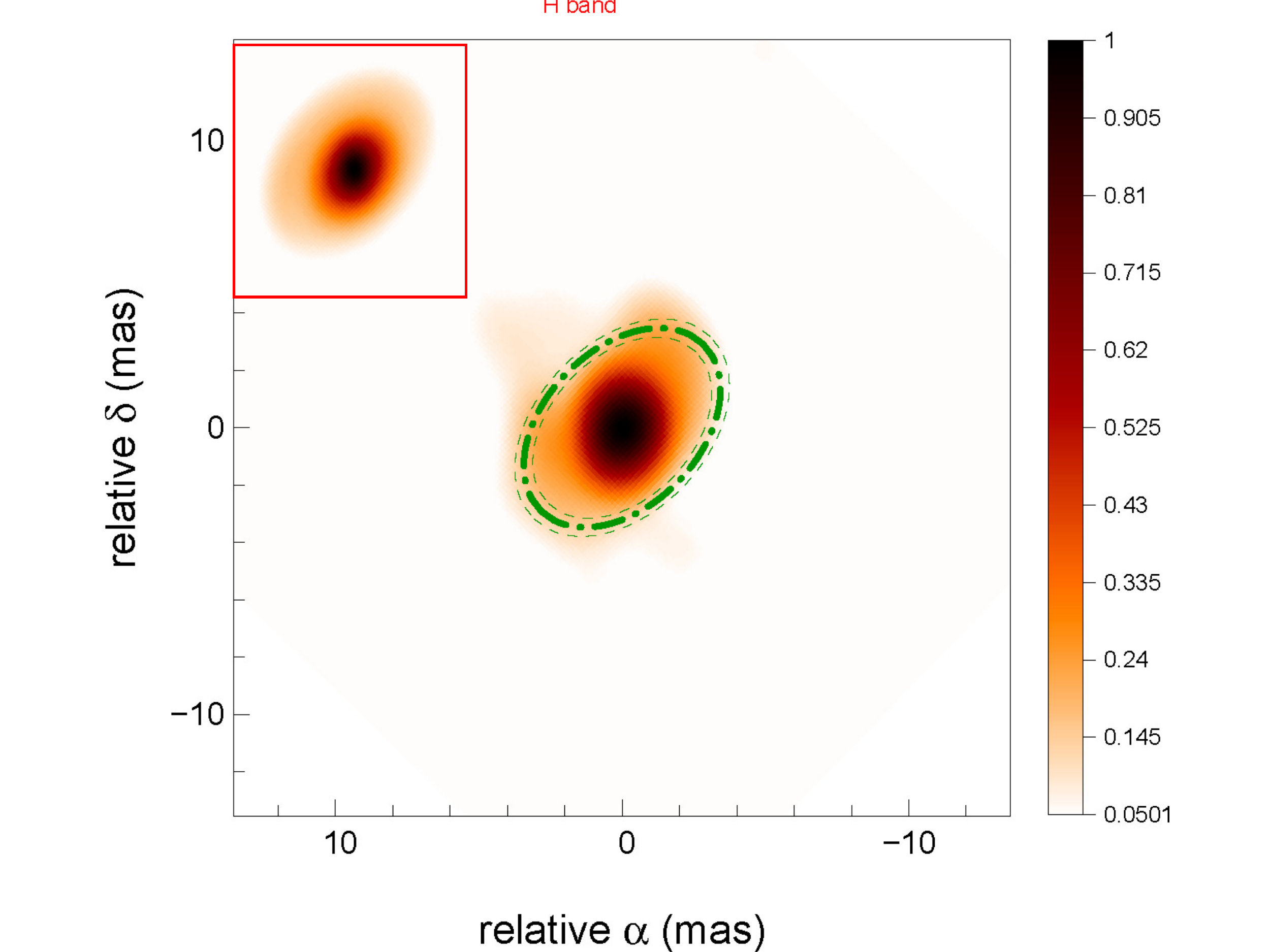}
  \hfill
  \includegraphics[height=6.7cm]{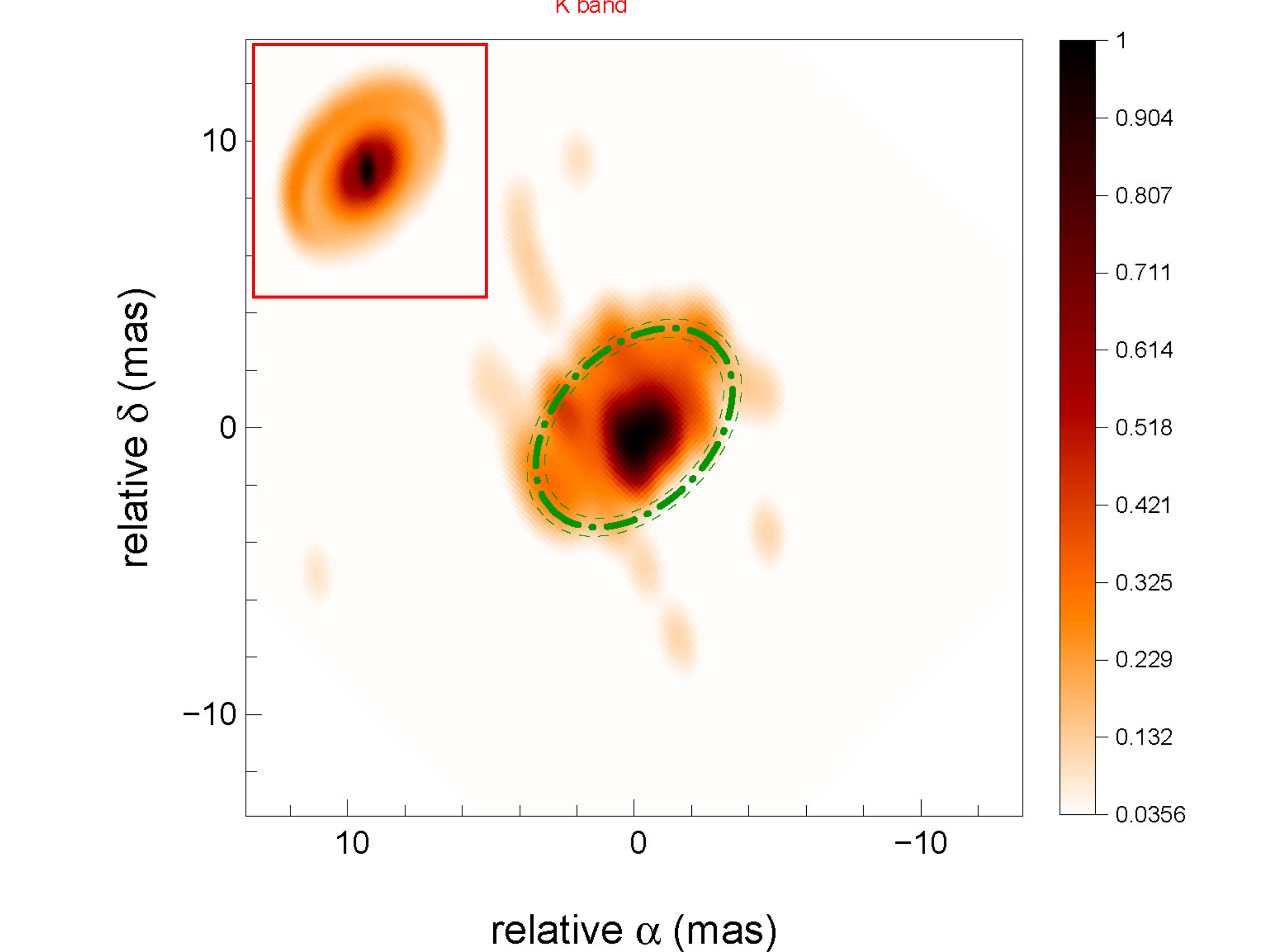}
  \hfill ~
  \caption{Reconstructed images of the B10 model of \mwc\ in the $H$
    (left) and $K$ bands (right). The dashed green ellipse
    corresponds to the location of the rim in this model. The
    models used are presented in the upper left corner. Same
    conventions as in Fig.~\ref{fig:real-img-rec}.}
  \label{fig:model-img-rec}
\end{figure*}
\begin{figure*}
  \centering
  \hfill
  \includegraphics[height=6.7cm]{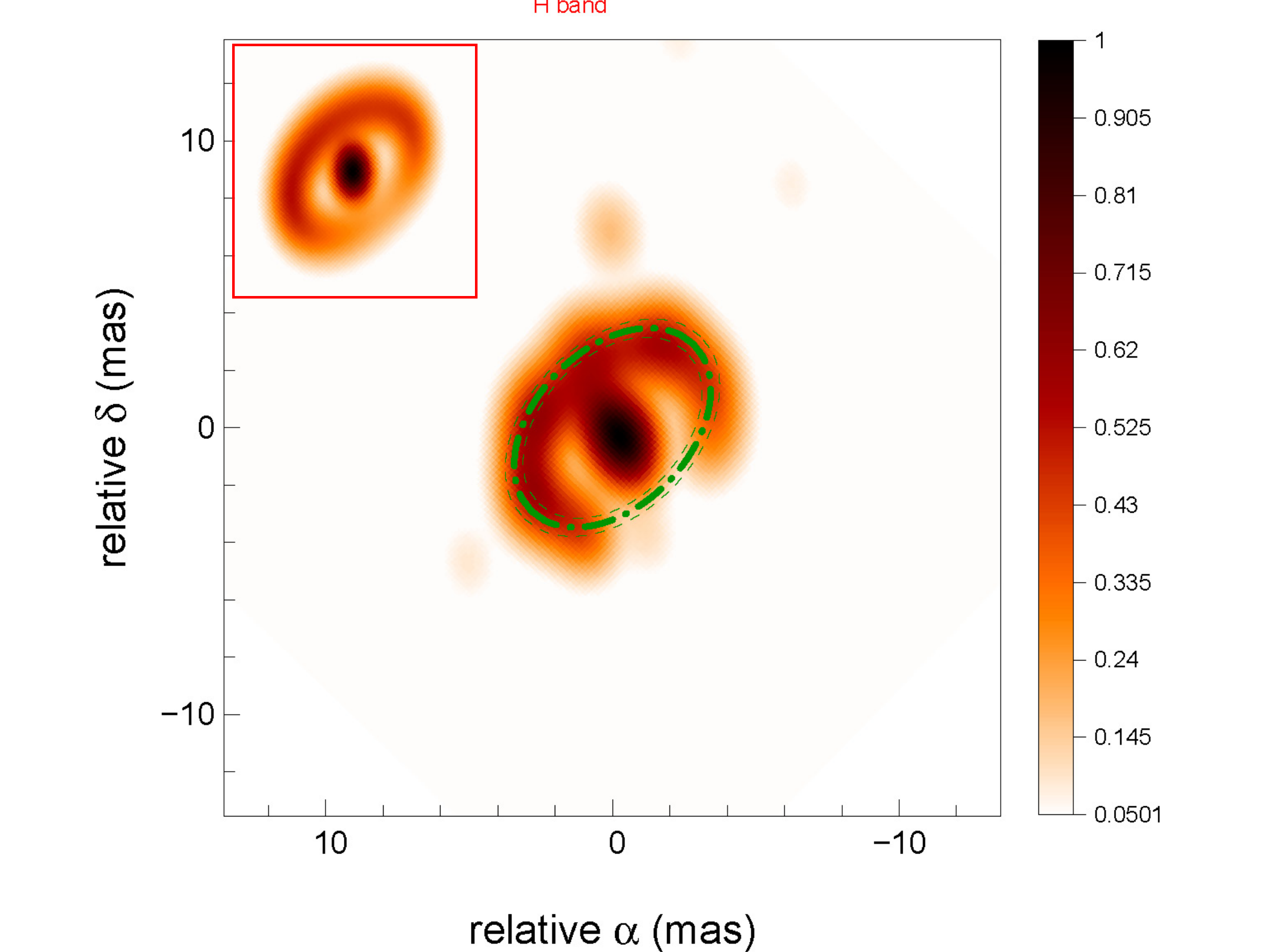}
  \hfill
  \includegraphics[height=6.7cm]{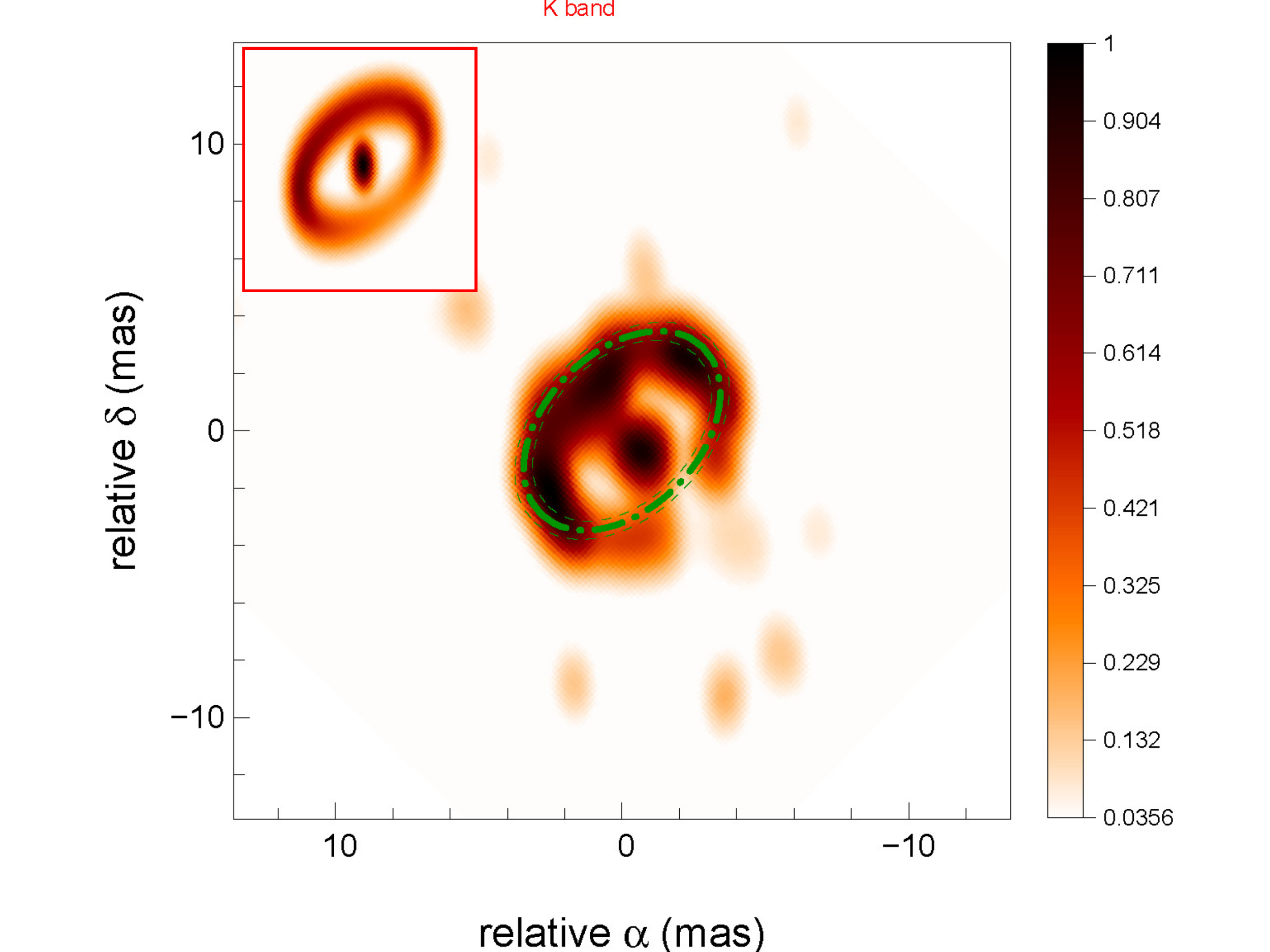}
 \hfill ~
 \caption{Reconstructed images of a geometrical model of \mwc\ with
   only a star plus a Gaussian ring in the $H$ (left) and $K$ bands
   (right). The dashed green ellipse corresponds to the location of
   the rim in this model. The models are presented in the upper left corner. Same conventions as in Fig.~\ref{fig:real-img-rec}.}
  \label{fig:modelSansInt-img-rec}
\end{figure*}

To analyze the artifacts in the images, we use
the model of \mwc\ presented in B10 (see the bottom right squares of
Fig.~\ref{fig:model-img-rec} for the $H$ and $K$-band models) to
simulate images with the same conditions as the actual ones.
The model is composed of a star (producing 30~\% and
14~\% of the flux in the $H$ and $K$ bands, respectively, estimated from
  spectral energy distribution fitting), a dust rim located
at 0.45\,AU ($\sim$3.6~mas) representing 16~\% (36~\%) in the $H$ ($K$) band
and a bright inner disk, from 0.1 to 0.45\,AU, contributes to the remaining emission. The reconstructed images of the model from simulated
data with the same \uv coverage and the same error bars as the real
data are shown in Fig.~\ref{fig:model-img-rec} for the $H$ and $K$ bands.

The analysis of the reconstructed images of the model in the $K$ band
illustrates that the following structures are well retrieved by the image
reconstruction process:
\begin{itemize}
\item The dust rim, which is clearly visible at the right location and
  appears  as a  somewhat  blobby  ellipse. We  checked that  by
  changing the \uv plane filling, the blob location changes but remains
  aligned along the ellipse.
\item The energy inside the disk, which is more spread than if there was only
  the star inside the dust rim in the model (see below). This emission between the
  star and the rim represents the bright inner disk.
\item The skewness of the dust rim is visible in the reconstructed
  images: the bottom part of the blobby ellipse has less flux than
  the top part.
\item The star, the dust rim, and the inner disk provide $\sim$15\%, $\sim$30\%, and $\sim$55\%
  of the flux, respectively. We emphasize that these values are in close agreement with the model. 
\end{itemize}
The only structure that is not retrieved in the image is the hole
between the star and the bright inner disk, inside 0.1\,AU, which is too
small to be resolved by the interferometer.

In the $H$ band, the dust rim   disappears in the reconstructed
image as a ring of blobs, but still seems to define the outer boundary
of the object. The main reasons are that the rim represents
only 16\% of the flux to be compared with 36\% in the $K$ band, and,
that the angular resolution is not high enough in the $H$ band
data compared to the $K$ band data (with the CHARA very long baselines). For the dust rim
to be seen  in the $H$ band, we would need  data on longer baselines
and higher dynamics in the reconstructed image (1000 at least, 2000 to be
unambiguously seen). As in the $K$ band, the bright inner disk is also
present in the $H$ band image as a large spot in the middle, which  would not have existed with the star only (see below), and
  represents 86\% of the total flux.

To demonstrate that the bright inner disk is clearly seen in the
reconstructed images, \ie, that the central spot includes more energy
than that from the star alone, reconstruction of a simpler model is
performed. This model is the same as the B10 model but without the bright
inner disk, \ie a star surrounded by a Gaussian ring.
The  star fluxes in  the $H$ and $K$  bands remain the same  and the
Gaussian ring accounts for 70\% of the flux in the $H$ band and 86\% in the $K$
band. Figure~\ref{fig:modelSansInt-img-rec} clearly indicates
that the star alone does not spread across more than over 4
pixels in the $K$ band and 7 in the $H$ band, which is less than in Fig.~\ref{fig:model-img-rec}. \newline

This analysis performed on existing models allows us to state which
features in the reconstructed images from Fig.~\ref{fig:real-img-rec} can be trusted. We argue that
the main secondary blobs present around the main central spot are
real. Their spatial distribution along an ellipse and the intensity
present between these peaks and the central spot are also real. However, the clumpy structure of the ring is probably not representative of the
reality, but only of the actual \uv plane. More observations at
different spatial frequencies will probably change the actual position
of these peaks along the ellipse, which may be smoothed. However, we
conclude that the inclination and orientation of the observed
distribution of peaks along an ellipse are real. This
orientation and inclination are indeed very close to the ones fitted by
B10 and \citet{Tannirkulam2008}, and are consistent with previous estimates
at different wavelengths \citep{Isella2007}. The second feature that we think is representative of the reconstructed image is that the central spot is extended and not
reduced to an unresolved point as a point-like star would be. The shape of
this central spot is certainly dependent on the filling of the \uv plane, although the position of the centroid is certainly
representative  of  reality.


\section{Discussion}
\label{sec:discussion}

In this section,  we discuss the reconstructed images.

\subsection{Dynamic range}
\label{sec:dynamics}

To  compute the theoretical  dynamical range  of our  image, we  use an
estimator based on the one proposed by \citet{Baldwin2002}
\begin{equation}
\sqrt{ \frac{n}{\left( \delta V/V \right)^2 + \left( \delta \mathrm{CP} \right)^2} } \mbox{\,\,\, ,}
\label{eq:dyn}
\end{equation}
where $n$ is the total number of measurements, $\delta V/V$ is the relative error in
the visibilities, and $\delta \mathrm{CP}$ is the error in the
closure phases (in radian). This number indicates the maximum contrast
that can be reached in the image given the data, \ie the ratio of
the maximum of the image to the minimum value that can be trusted.

Applying Eq.~\eqref{eq:dyn} to the data, a dynamic range of 780 is
found in the $K$ band. In the $H$ band, for which there is fewer data and the error
bars are slightly larger, a dynamic range of 400 is computed. Minimum
cut levels of 1/780 and 1/400, respectively,  are applied to all the $K$ and $H$ figures.

\subsection{Use of the spectral information}
\label{sec:wavelength}

\begin{figure}
  \resizebox{\hsize}{!}{
    \includegraphics{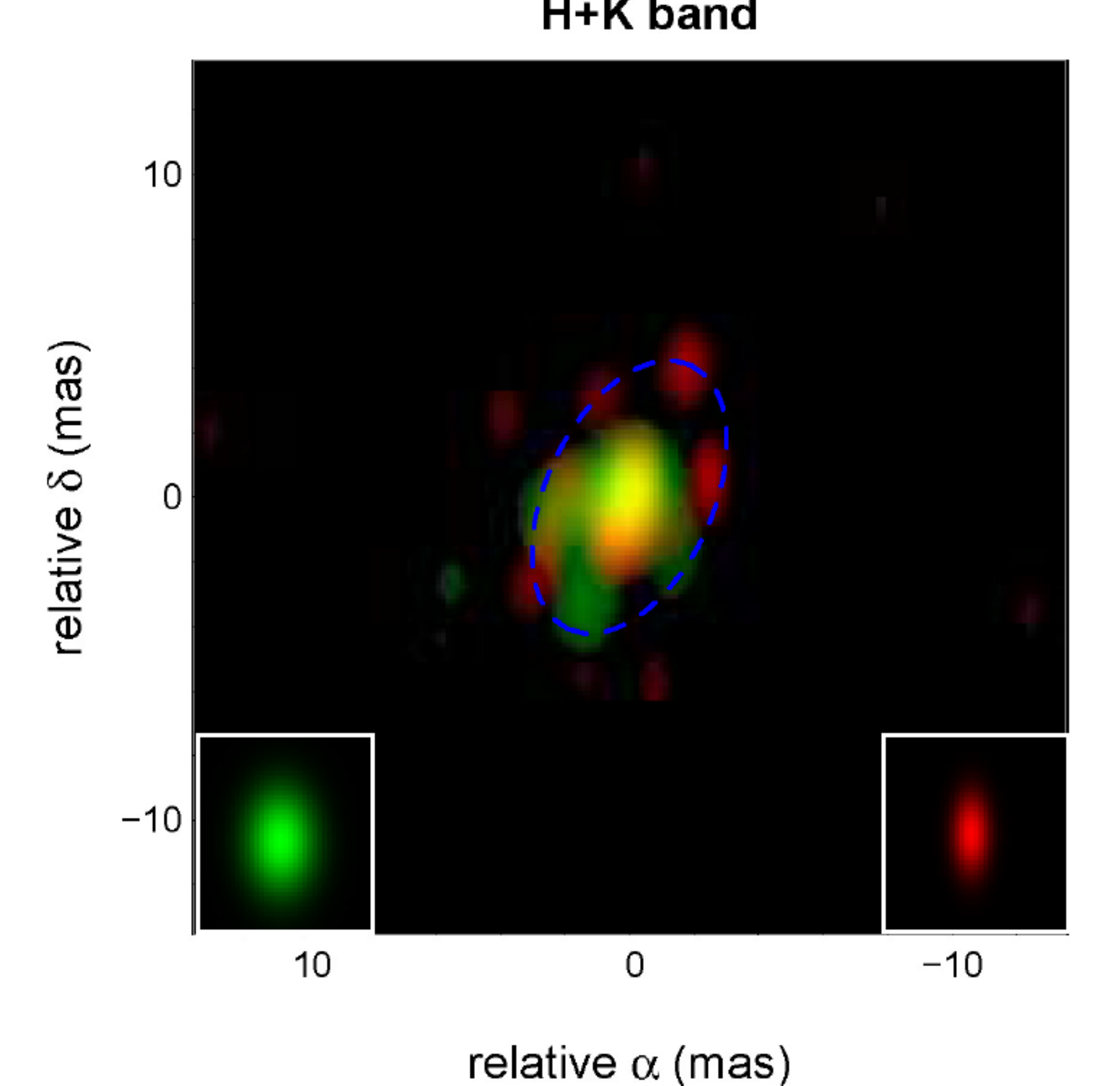}}
  \caption{Combination of the reconstructed images of \mwc\ in a two-color image
  ($H$ band in green, $K$ band in red). The blue ellipse traces the location of the main secondary blobs of the K-band emission. The sub-panels indicate the Gaussian beam at the
   interferometer resolution used in the convolution.}
  \label{fig:2colors}
\end{figure}

The decision to present one reconstructed image in the $H$ and one in the $K$
band results from different tests made on the B10 model. A trade-off
has to be made between ensuring that we have enough data to pave the \uv plane and the
wavelength dependency expected from circumstellar disks. Because of
the intrinsic chromaticity of the object, we prefer to reconstruct
two separated images in the $H$ and $K$ bands, otherwise two separated
visibilities, sampling different emitting regions, may correspond to the same spatial frequency. We show in Fig.~\ref{fig:2colors} the combination of both reconstructed images that illustrate complementary features.

For each band, we used all the data points in the different spectral channels,
assuming implicitly the object to be grey in each separated
band. Compared to the model, this method produces images with fewer
artifacts than when reconstructing a single broadband image with the average
data for all spectral channels. The \uv plane is indeed far more
filled taking into account all the wavelengths, because there are more
spatial frequencies.

\subsection{Physical consequences for the models}
\label{sec:discussion-model}

We emphasize  that we have obtained a new type of data in the form of images reconstructed with limited assumptions. The analysis of the results described in
Sect.~\ref{sec:analysis} is useful for distinguishing artifacts caused mainly by
the shape of the \uv plane from what we infer to be \emph{true}
features. In this section, we highlight the new information
provided by these images, but are also aware that these new pieces of
evidence have to be handled with great care.

The ellipse described by the successive blobs in the reconstructed
image at a distance of 4-5\,mas from the center of the image and
underlined by a blue dot-dashed line in Figs.~\ref{fig:real-img-rec} and
\ref{fig:contour-img-rec} certainly traces an external ring. The
characteristics of this ring are not exactly the same as the rim found
in the B10 model plotted with a green dashed line in
Figs.~\ref{fig:real-img-rec} and \ref{fig:contour-img-rec}. The
radius of the ring is 0.55\,AU instead of $0.45 \pm 0.05$\,AU in the
B10 model, the inclination is 55\degr\ instead of $48 \degr\ \pm 2\degr$, and the position
angle is 155\degr\ instead of $136 \degr\ \pm 2\degr$. The intensity is close to that of the model with 30\% of the total flux instead of 36\%. The location of the blobs
differs a little bit between the $H$ and $K$ bands: those in the $H$-band image are closer to the center of the image than those in
the $K$-band image. This behavior could be explained by a temperature gradient in the disk,  which has a tendency to move the
peak of intensity closer to the center for the shorter wavelengths. We
probably also need data on longer baselines (equivalent to the $K$-band CHARA data) to resolve unambiguously the external ring in the $H$-band
image.  This bright ring is not clear evidence of a physical rim, which  was proposed by
\citet{Dullemond2001}. We propose that this feature instead traces an
enhancement of the intensity caused by a change in the opacity in the
disk probably due to sublimation of dust.

We assume that the number of blobs along this ellipse is representative
of the azimuthal distribution of the intensity. Following B10 who in their analysis proposed  a model of a rim with a
skewed distribution of intensity along the ring (see
Figs.~\ref{fig:modelSansInt-img-rec}), we assume that the distribution
of light along the blue ellipse indicates that the actual ring of light is
less luminous in the south-west direction than in the north-east direction at
least in the $K$-band image. In the $H$-band image, there are blobs
around the central spot that do not appear in the reconstructed
image of the model. Does this mean that there is more flux in the
external ring than expected? Is the skewness factor more important
because the blobs appeared at one side only? Does the presence of blobs
in  the south part  of the  ellipse rather  than in  the north  in the
$K$-band image have some significance? To remain on solid ground, we assume
that the intensity varies along the distribution of blobs, but do
not have definitive data to determine the magnitude of this
effect.  The departure  from  the axisymmetry may be caused by  an
inclined surface of the disk or even a strong dust opacity change.

The energy detected within the ellipse is certainly real: as explained in
Sect.~\ref{sec:analysis}, the central region, which contains about 70\%
of the flux, certainly does not originate in a single unresolved star but
from an extended source. These results independently confirm the conclusions of B10, without using any model.
Indeed, B10 were unable to fit the visibilities at higher spatial frequencies without
introducing some continuous emission in the space between the rim and
the star. They called this region the inner disk.  We do not know whether the shape of this central source
in the $K$-band image in the form of two spots is real or not, since in
the simulated image this central source also seems to be decomposed into
3 single sources, which were not present in the model. It might be only the
effect of the \uv plane coverage, but we cannot exclude too that it
might be due to a hot spot in the disk, although we should in
principle then see it in the two images. The reconstructed images cannot help us to determine the origin of the inner disk emission, and ascertain whether it comes from hot gas or very refractory grains. These unknowns may be solved by combining high resolution spectroscopic observations in the near-infrared with advanced models that self-consistently compute the emission of both dust and gas.

Finally, it remains unclear whether the non-zero value of the closure phases found by
B10 is caused by the contribution from the inner
  disk not being exactly centered with the external ring. This might be the case when the disk
surface is flaring and the system is seen with non-zero inclination. If
the curvature of the surface of the disk probed by our images is large
enough, then the ellipses tracing equal distances to the star will be
shifted in the polar direction.

\subsection{Consequences on the image reconstruction}
\label{sec:discussion-methods}

Since image reconstruction in optical interferometry remains in
  its  infancy, only a too sparse \uv coverage is  available  to reconstruct  an
unambiguous image of a complex object and analyze it without the help of the model-fitting
technique. Repeated comparisons between the model and the reconstructed
image have to be performed to avoid over-interpreting the structures in
the images. However, in all these cases, the image reconstruction
technique remains the only technique able to perform
model-independent analysis of the data and is a powerful tool to give more credits to the models, derive tighter constraints, or even reveal unexpected structures.

In the future, two aspects of the image reconstruction will be important:
\begin{enumerate}
\item \textbf{The homogeneity of the \uv plane}: thanks to several
  tests on the B10 model, we  found that the global blobby aspect
  of  the reconstructed image is almost certainly caused by the non-homogeneity of the
  \uv plane. Indeed the reconstructed images from a homogeneous \uv
  plane are smoother. The holes in the \uv plane correspond to blobby or
  point-like structures in the reconstructed images, and the
  baselines for the observations have to be carefully chosen to map
  the \uv plane as homogeneously as possible.
\item \textbf{The quality of the measurements}: the number of data points and their associated errors can clearly determine the dynamic range of the
  image. A larger number of  data points and  smaller error bars
  are needed to improve the dynamics of the image.
\end{enumerate}
\begin{figure}
  \resizebox{\hsize}{!}{
    \includegraphics{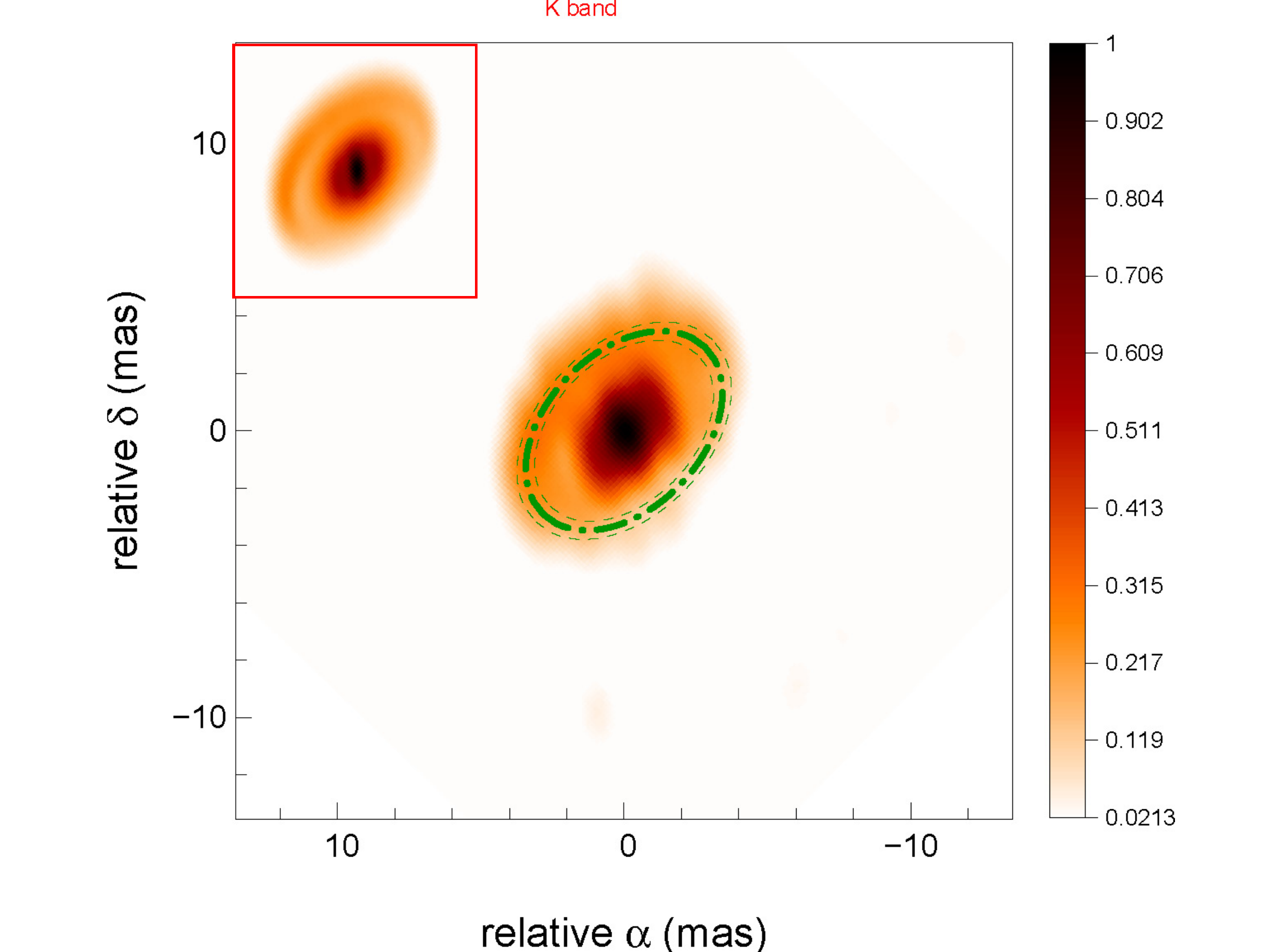}}
  \caption{Reconstructed image of the B10 model in the $K$ band, using a
    synthetic \uv plane obtained with 3  quadruplets (A0-K0-G0-I1,
    D0-H0-G1-I1, E0-G0-H0-I1) at VLTI  and the CHARA baselines (S1-W1,
    W1-W2, S2-W2, E1-W1, E2-S2, S2-W1).}
  \label{fig:pionier}
\end{figure}
These aspects are illustrated in Fig.~\ref{fig:pionier} where data on the B10 model are simulated with a
\uv plane for 3 quadruplets of telescopes at VLTI, as offered
in the coming period P86 for the AMBER instrument and later for the
PIONIER visitor instrument. The
reconstructed image  from these  data contains fewer  artifacts than
those from  the actual data set.   The advent  of the
next generation  imaging instruments at  the VLTI, such  as GRAVITY,
MATISSE, and VSI, with up to  8 telescopes, will allow us to study in
detail the inner disk structure, to detect planetary gaps and map
outflows  -- bringing unambiguous  constraints on  crucial mechanisms
for star and planet formation.


\section{Conclusion}
\label{sec:conclusion}

Since the renewal of optical interferometry in the mid-70's, imaging
using aperture synthesis between different telescopes has been
attempted for all possible astrophysical topics following the path
opened by radio-interferometry with the Very Large
Array. Unfortunately, optical interferometry requires precisions in the optical path delay
$10^3$ to $10^5$ smaller than in the radio domain and therefore the
development of large optical arrays has been slower. Although images of stellar surfaces or binaries have been obtained, no synthetic
images have so far been possible at the milli-arcsecond scale of complex young
stellar objects.

This paper is the first one attempting to reconstruct images of the
close environment of a young star independent of any a priori
model by gathering many interferometric measurements on the young star
\mwc\ with various interferometers (VLTI, IOTA, KeckI, and
CHARA). Using the Multi-Aperture Image Reconstruction Algorithm
(\Mira), we have reconstructed images of \mwc\ in the $H$ and $K$ bands. The
quality of the images compares well to those obtained 40 years ago
with the first three antennas of the Very Large Array.  To
assess the reality of the features that are present, we have compared these
images with images obtained from simulated data using the best known
physical model of the environment of \mwc\ and reconstructed with the
same conditions of \uv plane and noise.

Those model-independent $H$ and $K$-band images of the
surrounding of \mwc\ reveal several significant features that we can
relate to the presence of an inclined asymmetric disk around \mwc.
We also detect a pre-eminent intensity at the location of  dust sublimation
above a structure extending from the central source with some
differential effect  along the azimuth  that is not necessarily
related to a puffed-up inner rim. Together  with a slight
offset of the central source compared to this bright ring, we
proposed this to be the signature of the flared surface of a disk with a
discontinuity caused by a opacity change.

These images confirm that the morphology of the close environment of
young stars is more complex than the simple models used in
the literature. We have also shown that having a more uniform \uv plane
coverage and higher measurement accuracies should help us to obtain higher quality images in the future.


\begin{acknowledgements}
  The authors are grateful to the different institutions that trusted
  them in giving them guaranteed observing time with the VLTI to
  perform the first image of a complex young stellar object: CNRS through its
  national programs (ASHRA and PNPS), INAF and the AMBER
  consortium.  This research has made use of NASA's Astrophysics Data
  System service, of the Jean-Marie Mariotti Center (JMMC) resources
  of the SIMBAD database, operated at CDS, Strasbourg, France, and of
  the Yorick, a free data processing language written by D.\ Munro (\url{http://yorick.sourceforge.net}).
  M.B. acknowledges fundings from INAF (grant ASI-INAF I/016/07/0).
\end{acknowledgements}


\bibliographystyle{aa} 
\bibliography{srenard_14910_accepted_final}

\begin{thebibliography}{39}
\expandafter\ifx\csname natexlab\endcsname\relax\def\natexlab#1{#1}\fi

\bibitem[{{Balbus} \& {Hawley}(1991)}]{Balbus1991}
{Balbus}, S.~A. \& {Hawley}, J.~F. 1991, \apj, 376, 214

\bibitem[{{Baldwin} \& {Haniff}(2002)}]{Baldwin2002}
{Baldwin}, J.~E. \& {Haniff}, C.~A. 2002, Royal Society of London Philosophical
  Transactions Series A, 360, 969

\bibitem[{{Benisty} {et~al.}(2010){Benisty}, {Natta}, {Isella}, {Berger},
  {Massi}, {Le Bouquin}, {M{\'e}rand}, {Duvert}, {Kraus}, {Malbet}, {Olofsson},
  {Robbe-Dubois}, {Testi}, {Vannier}, \& {Weigelt}}]{Benisty2010}
{Benisty}, M., {Natta}, A., {Isella}, A., {et~al.} 2010, \aap, 511, A74+

\bibitem[{{Boss}(1997)}]{Boss1997}
{Boss}, A.~P. 1997, Science, 276, 1836

\bibitem[{{Cotton} {et~al.}(2008){Cotton}, {Monnier}, {Baron}, {Hofmann},
  {Kraus}, {Weigelt}, {Rengaswamy}, {Thi{\'e}baut}, {Lawson}, {Jaffe},
  {Hummel}, {Pauls}, {Schmitt}, {Tuthill}, \& {Young}}]{Cotton2008}
{Cotton}, W., {Monnier}, J., {Baron}, F., {et~al.} 2008, in Society of
  Photo-Optical Instrumentation Engineers (SPIE) Conference Series, Vol. 7013,
  Society of Photo-Optical Instrumentation Engineers (SPIE) Conference Series

\bibitem[{{Deleuil} {et~al.}(2005){Deleuil}, {Bouret}, {Catala}, {Lecavelier
  des Etangs}, {Vidal-Madjar}, {Roberge}, {Feldman}, {Martin}, \&
  {Ferlet}}]{Deleuil2005}
{Deleuil}, M., {Bouret}, J., {Catala}, C., {et~al.} 2005, \aap, 429, 247

\bibitem[{{Devine} {et~al.}(2000){Devine}, {Grady}, {Kimble}, {Woodgate},
  {Bruhweiler}, {Boggess}, {Linsky}, \& {Clampin}}]{Devine2000}
{Devine}, D., {Grady}, C.~A., {Kimble}, R.~A., {et~al.} 2000, \apjl, 542, L115

\bibitem[{{Dullemond} {et~al.}(2001){Dullemond}, {Dominik}, \&
  {Natta}}]{Dullemond2001}
{Dullemond}, C.~P., {Dominik}, C., \& {Natta}, A. 2001, \apj, 560, 957

\bibitem[{{Grady} {et~al.}(2000){Grady}, {Devine}, {Woodgate}, {Kimble},
  {Bruhweiler}, {Boggess}, {Linsky}, {Plait}, {Clampin}, \&
  {Kalas}}]{Grady2000}
{Grady}, C.~A., {Devine}, D., {Woodgate}, B., {et~al.} 2000, \apj, 544, 895

\bibitem[{{Haubois} {et~al.}(2009){Haubois}, {Perrin}, {Lacour}, {Verhoelst},
  {Meimon}, {Mugnier}, {Thi{\'e}baut}, {Berger}, {Ridgway}, {Monnier},
  {Millan-Gabet}, \& {Traub}}]{Hautbois2009}
{Haubois}, X., {Perrin}, G., {Lacour}, S., {et~al.} 2009, \aap, 508, 923

\bibitem[{{Henning} \& {Meeus}(2009)}]{HenningMeeus2009}
{Henning}, T. \& {Meeus}, G. 2009, in "Physical Processes in Circumstellar
  Disks around Young Stars", Garcia, PJV (Ed.), Theoretical Astrophysics
  Series, Chicago University Press, Vol. in press (arXiv:0911.1010)

\bibitem[{{Hogg} {et~al.}(1969){Hogg}, {MacDonald}, {Conway}, \&
  {Wade}}]{Hogg1969}
{Hogg}, D.~E., {MacDonald}, G.~H., {Conway}, R.~G., \& {Wade}, C.~M. 1969, \aj,
  74, 1206

\bibitem[{{Isella} {et~al.}(2007){Isella}, {Testi}, {Natta}, {Neri}, {Wilner},
  \& {Qi}}]{Isella2007}
{Isella}, A., {Testi}, L., {Natta}, A., {et~al.} 2007, \aap, 469, 213

\bibitem[{{Kraus} {et~al.}(2009){Kraus}, {Weigelt}, {Balega}, {Docobo},
  {Hofmann}, {Preibisch}, {Schertl}, {Tamazian}, {Driebe}, {Ohnaka}, {Petrov},
  {Sch{\"o}ller}, \& {Smith}}]{Kraus2009}
{Kraus}, S., {Weigelt}, G., {Balega}, Y.~Y., {et~al.} 2009, \aap, 497, 195

\bibitem[{{Lacour} {et~al.}(2008){Lacour}, {Meimon}, {Thi{\'e}baut}, {Perrin},
  {Verhoelst}, {Pedretti}, {Schuller}, {Mugnier}, {Monnier}, {Berger},
  {Haubois}, {Poncelet}, {Le Besnerais}, {Eriksson}, {Millan-Gabet}, {Ragland},
  {Lacasse}, \& {Traub}}]{Lacour2008}
{Lacour}, S., {Meimon}, S., {Thi{\'e}baut}, E., {et~al.} 2008, \aap, 485, 561

\bibitem[{Lacour {et~al.}(2009)Lacour, Thiébaut, Perrin, Meimon, Haubois,
  Pedretti, Ridgway, Monnier, Berger, Schuller, Woodruff, Poncelet, Coroller,
  Millan-Gabet, Lacasse, \& Traub}]{Lacour_et_al-2009-Chi_Cygni}
Lacour, S., Thiébaut, E., Perrin, G., {et~al.} 2009, \apj, 707, 632

\bibitem[{{Lawson}(2000)}]{Lawson2000}
{Lawson}, P.~R., ed. 2000, {Principles of Long Baseline Stellar Interferometry}

\bibitem[{{Le Bouquin} {et~al.}(2009){Le Bouquin}, {Lacour}, {Renard},
  {Thi{\'e}baut}, {Merand}, \& {Verhoelst}}]{LeBouquin2009}
{Le Bouquin}, J., {Lacour}, S., {Renard}, S., {et~al.} 2009, \aap, 496, L1

\bibitem[{{Malbet} \& {Perrin}(2007)}]{Malbet2007}
{Malbet}, F. \& {Perrin}, G. 2007, New Astronomy Review, 51, 563

\bibitem[{{Mannings} \& {Sargent}(1997)}]{Mannings1997}
{Mannings}, V. \& {Sargent}, A.~I. 1997, \apj, 490, 792

\bibitem[{{Mayer} {et~al.}(2002){Mayer}, {Quinn}, {Wadsley}, \&
  {Stadel}}]{Mayer2002}
{Mayer}, L., {Quinn}, T., {Wadsley}, J., \& {Stadel}, J. 2002, Science, 298,
  1756

\bibitem[{{Millan-Gabet} {et~al.}(2007){Millan-Gabet}, {Malbet}, {Akeson},
  {Leinert}, {Monnier}, \& {Waters}}]{MillanGabet2007}
{Millan-Gabet}, R., {Malbet}, F., {Akeson}, R., {et~al.} 2007, Protostars and
  Planets V, 539

\bibitem[{{Monnier}(2003)}]{Monnier2003}
{Monnier}, J.~D. 2003, Reports on Progress in Physics, 66, 789

\bibitem[{{Monnier} {et~al.}(2006){Monnier}, {Berger}, {Millan-Gabet}, {Traub},
  {Schloerb}, {Pedretti}, {Benisty}, {Carleton}, {Haguenauer}, {Kern},
  {Labeye}, {Lacasse}, {Malbet}, {Perraut}, {Pearlman}, \&
  {Zhao}}]{Monnier2006}
{Monnier}, J.~D., {Berger}, J., {Millan-Gabet}, R., {et~al.} 2006, \apj, 647,
  444

\bibitem[{{Monnier} {et~al.}(2005){Monnier}, {Millan-Gabet}, {Billmeier},
  {Akeson}, {Wallace}, {Berger}, {Calvet}, {D'Alessio}, {Danchi}, {Hartmann},
  {Hillenbrand}, {Kuchner}, {Rajagopal}, {Traub}, {Tuthill}, {Boden}, {Booth},
  {Colavita}, {Gathright}, {Hrynevych}, {Le Mignant}, {Ligon}, {Neyman},
  {Swain}, {Thompson}, {Vasisht}, {Wizinowich}, {Beichman}, {Beletic},
  {Creech-Eakman}, {Koresko}, {Sargent}, {Shao}, \& {van Belle}}]{Monnier2005}
{Monnier}, J.~D., {Millan-Gabet}, R., {Billmeier}, R., {et~al.} 2005, \apj,
  624, 832

\bibitem[{{Monnier} {et~al.}(2007){Monnier}, {Zhao}, {Pedretti}, {Thureau},
  {Ireland}, {Muirhead}, {Berger}, {Millan-Gabet}, {Van Belle}, {ten
  Brummelaar}, {McAlister}, {Ridgway}, {Turner}, {Sturmann}, {Sturmann}, \&
  {Berger}}]{Monnier2007}
{Monnier}, J.~D., {Zhao}, M., {Pedretti}, E., {et~al.} 2007, Science, 317, 342

\bibitem[{{Montesinos} {et~al.}(2009){Montesinos}, {Eiroa}, {Mora}, \&
  {Mer{\'{\i}}n}}]{Montesinos2009}
{Montesinos}, B., {Eiroa}, C., {Mora}, A., \& {Mer{\'{\i}}n}, B. 2009, \aap,
  495, 901

\bibitem[{{Natta} {et~al.}(2004){Natta}, {Testi}, {Neri}, {Shepherd}, \&
  {Wilner}}]{Natta2004}
{Natta}, A., {Testi}, L., {Neri}, R., {Shepherd}, D.~S., \& {Wilner}, D.~J.
  2004, \aap, 416, 179

\bibitem[{Strong \& Chan(2003)}]{Strong_Chan-2003-total_variation}
Strong, D. \& Chan, T. 2003, Inverse Problems, 19, S165

\bibitem[{{Swartz} {et~al.}(2005){Swartz}, {Drake}, {Elsner}, {Ghosh}, {Grady},
  {Wassell}, {Woodgate}, \& {Kimble}}]{Swartz2005}
{Swartz}, D.~A., {Drake}, J.~J., {Elsner}, R.~F., {et~al.} 2005, \apj, 628, 811

\bibitem[{{Tannirkulam} {et~al.}(2008){Tannirkulam}, {Monnier}, {Millan-Gabet},
  {Harries}, {Pedretti}, {ten Brummelaar}, {McAlister}, {Turner}, {Sturmann},
  \& {Sturmann}}]{Tannirkulam2008}
{Tannirkulam}, A., {Monnier}, J.~D., {Millan-Gabet}, R., {et~al.} 2008, \apjl,
  677, L51

\bibitem[{{Thi\'ebaut}(2002)}]{Thiebaut2002}
{Thi\'ebaut}, E. 2002, in Presented at the Society of Photo-Optical
  Instrumentation Engineers (SPIE) Conference, Vol. 4847, Society of
  Photo-Optical Instrumentation Engineers (SPIE) Conference Series, ed.
  {J.-L.~Starck \& F.~D.~Murtagh}, 174--183

\bibitem[{{Thi{\'e}baut}(2005)}]{Thiebaut2005}
{Thi{\'e}baut}, E. 2005, in NATO ASIB Proc. 198: Optics in astrophysics, ed.
  {R.~Foy \& F.~C.~Foy}, 397--+

\bibitem[{{Thi{\'e}baut}(2008)}]{Thiebaut2008}
{Thi{\'e}baut}, E. 2008, in Presented at the Society of Photo-Optical
  Instrumentation Engineers (SPIE) Conference, Vol. 7013, Society of
  Photo-Optical Instrumentation Engineers (SPIE) Conference Series

\bibitem[{{Thi{\'e}baut} \& {Giovannelli}(2009)}]{Thiebaut2009}
{Thi{\'e}baut}, {\'E}. \& {Giovannelli}, J.-F. 2009, ArXiv e-prints

\bibitem[{{van den Ancker} {et~al.}(1998){van den Ancker}, {de Winter}, \&
  {Tjin A Djie}}]{vandenAncker1998}
{van den Ancker}, M.~E., {de Winter}, D., \& {Tjin A Djie}, H.~R.~E. 1998,
  \aap, 330, 145

\bibitem[{{Wassell} {et~al.}(2006){Wassell}, {Grady}, {Woodgate}, {Kimble}, \&
  {Bruhweiler}}]{Wassell2006}
{Wassell}, E.~J., {Grady}, C.~A., {Woodgate}, B., {Kimble}, R.~A., \&
  {Bruhweiler}, F.~C. 2006, \apj, 650, 985

\bibitem[{{Zhao} {et~al.}(2008){Zhao}, {Gies}, {Monnier}, {Thureau},
  {Pedretti}, {Baron}, {Merand}, {ten Brummelaar}, {McAlister}, {Ridgway},
  {Turner}, {Sturmann}, {Sturmann}, {Farrington}, \& {Goldfinger}}]{Zhao2008}
{Zhao}, M., {Gies}, D., {Monnier}, J.~D., {et~al.} 2008, \apjl, 684, L95

\bibitem[{{Zhao} {et~al.}(2009){Zhao}, {Monnier}, {Pedretti}, {Thureau},
  {M{\'e}rand}, {ten Brummelaar}, {McAlister}, {Ridgway}, {Turner}, {Sturmann},
  {Sturmann}, {Goldfinger}, \& {Farrington}}]{Zhao2009}
{Zhao}, M., {Monnier}, J.~D., {Pedretti}, E., {et~al.} 2009, \apj, 701, 209

\end{thebibliography}


\clearpage
\appendix


\section{The data set in detail}
\label{app:data}

\begin{figure*}[b]
  \centering
  \begin{tabular}{cc}
    \includegraphics[width=6.5cm]{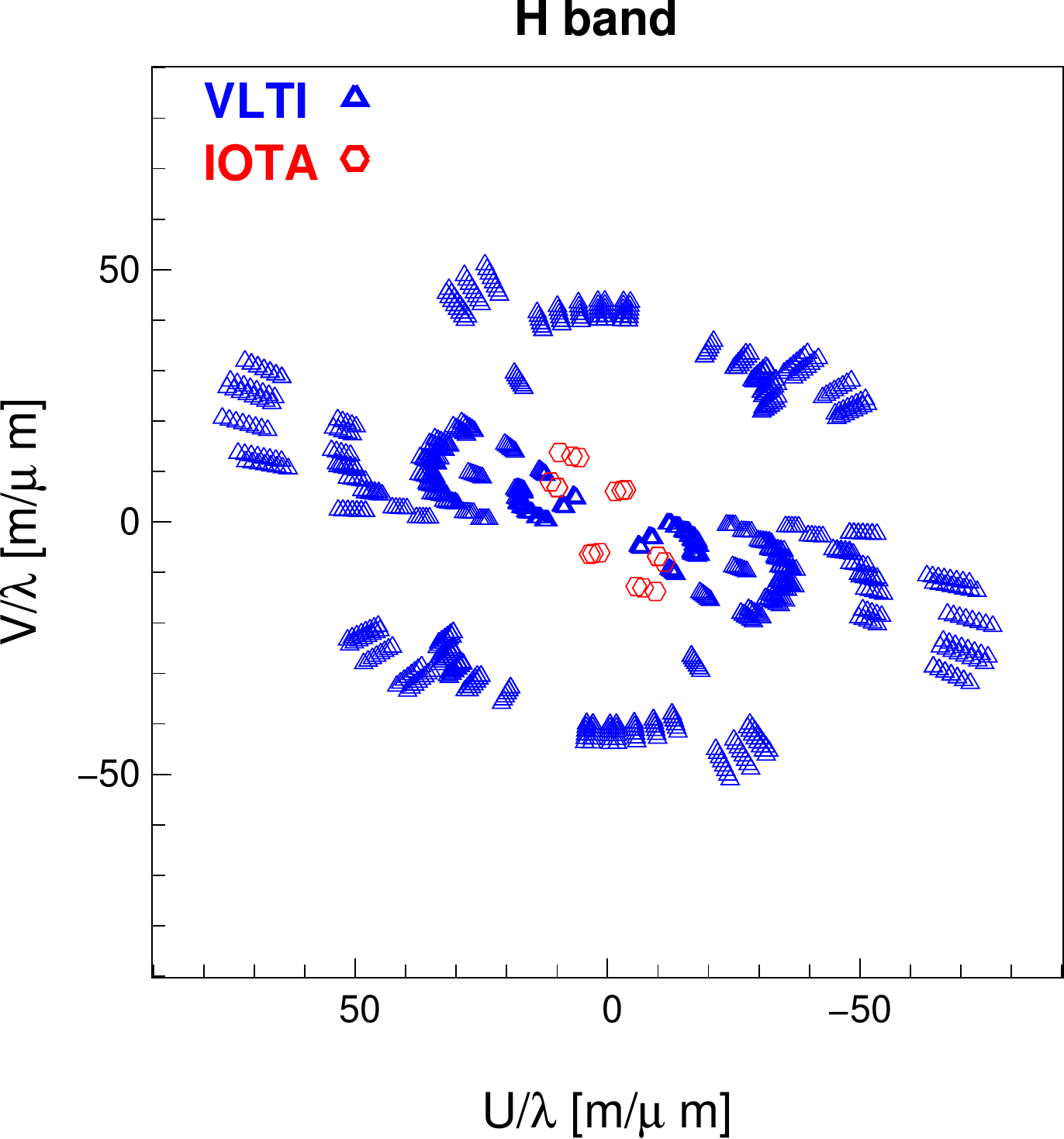}&
    \includegraphics[width=6.5cm]{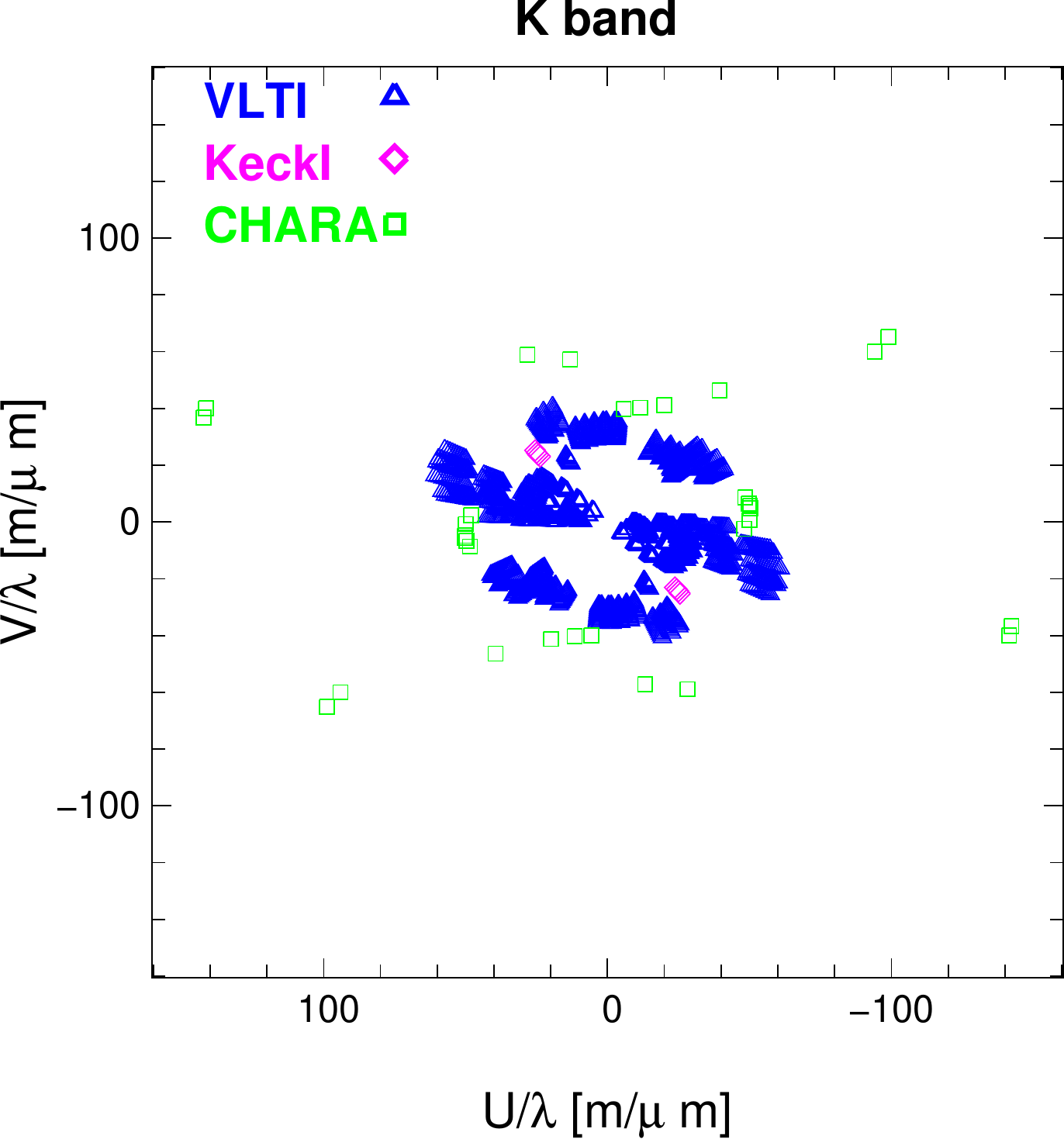}
  \end{tabular}
   \caption{\uv plane coverage of the data used for the image
     reconstruction in spatial frequencies for the $H$ (left) and $K$ (right) bands. The different
     interferometers are plotted in different colors and symbols.}
  \label{fig:uvcov}
\end{figure*}
\begin{figure*}[p]
  \centering
  \begin{tabular}{cc}
    \includegraphics[width=6.5cm]{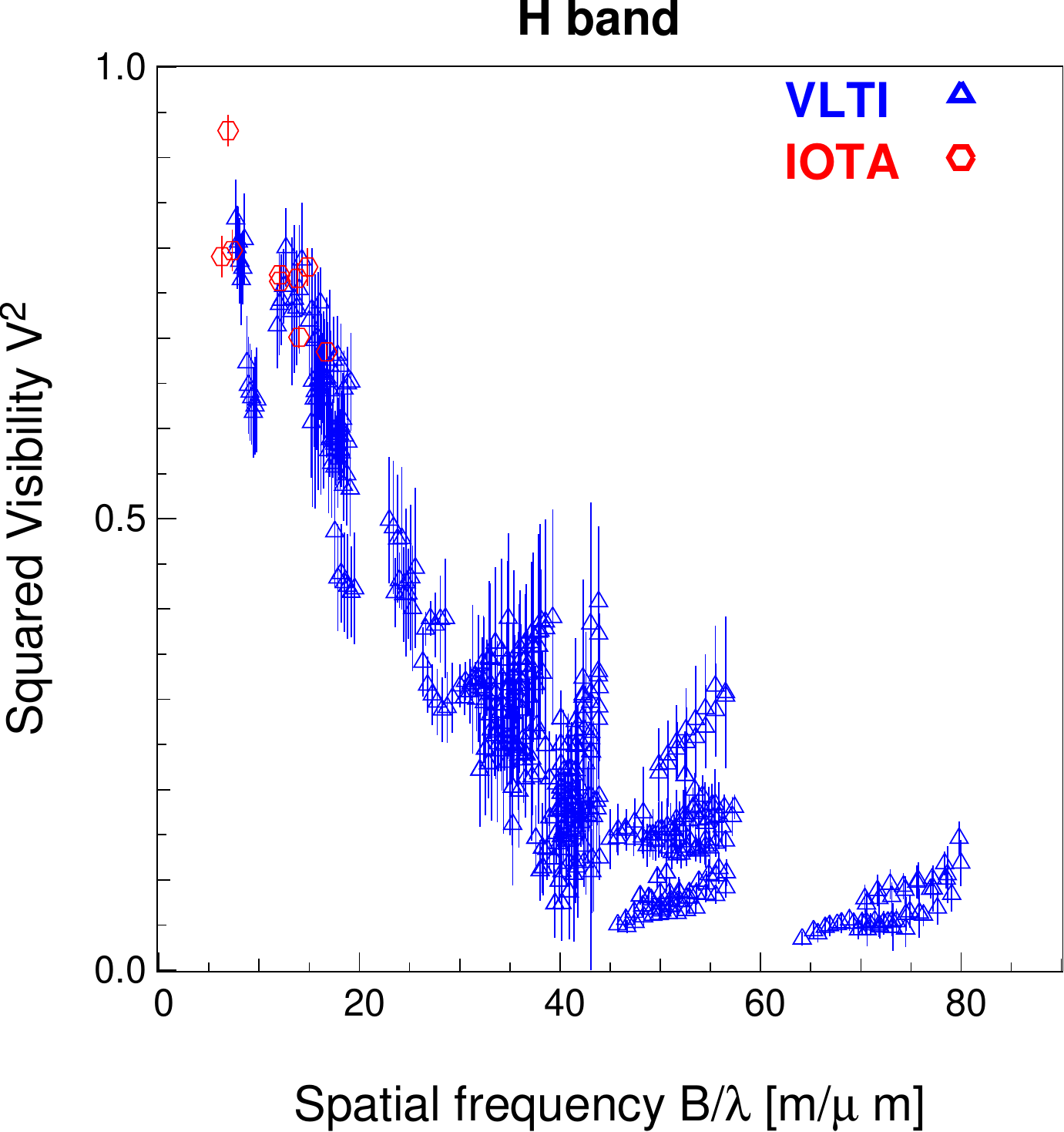}&
    \includegraphics[width=6.5cm]{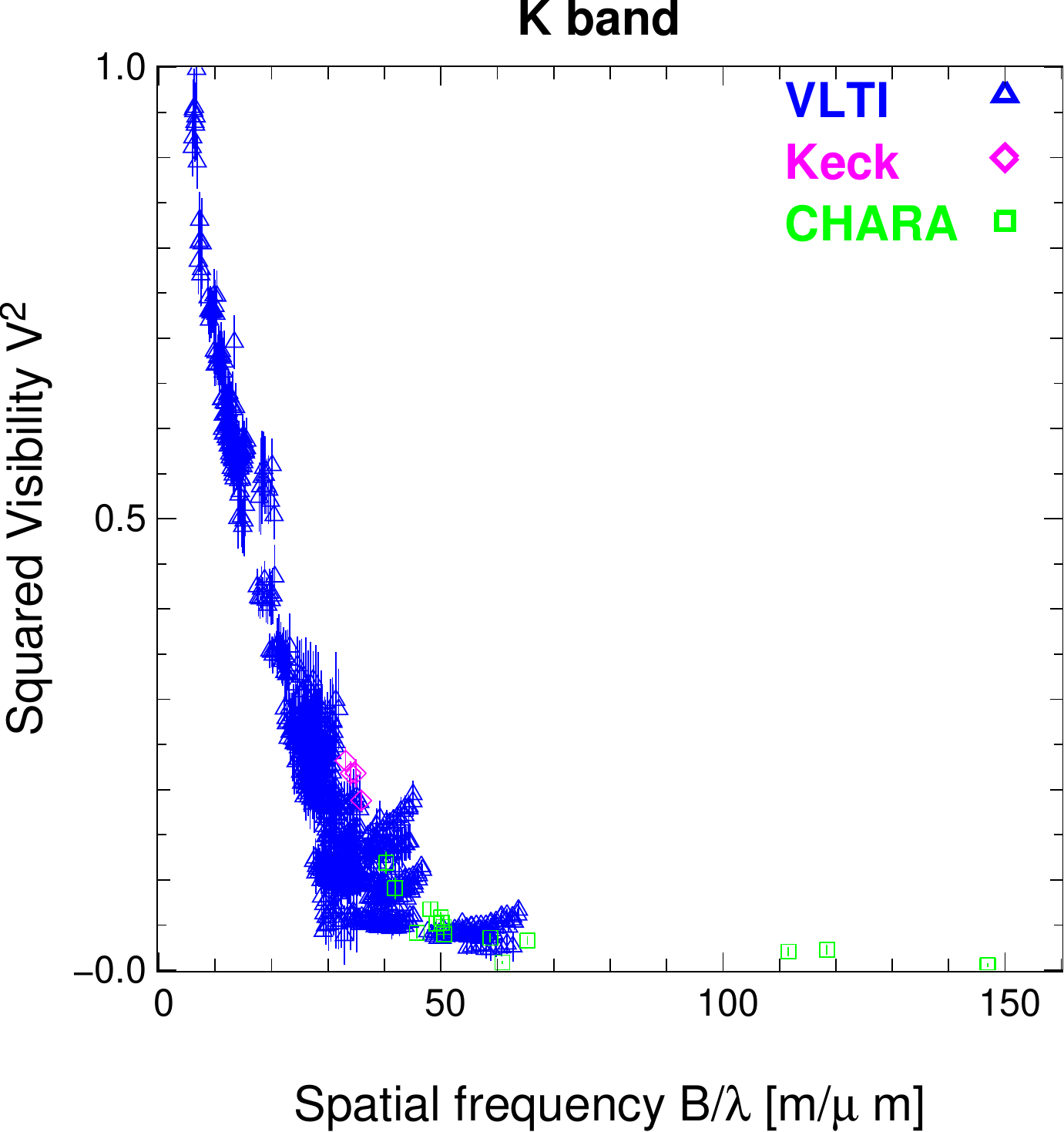}\\
    \includegraphics[width=6.5cm]{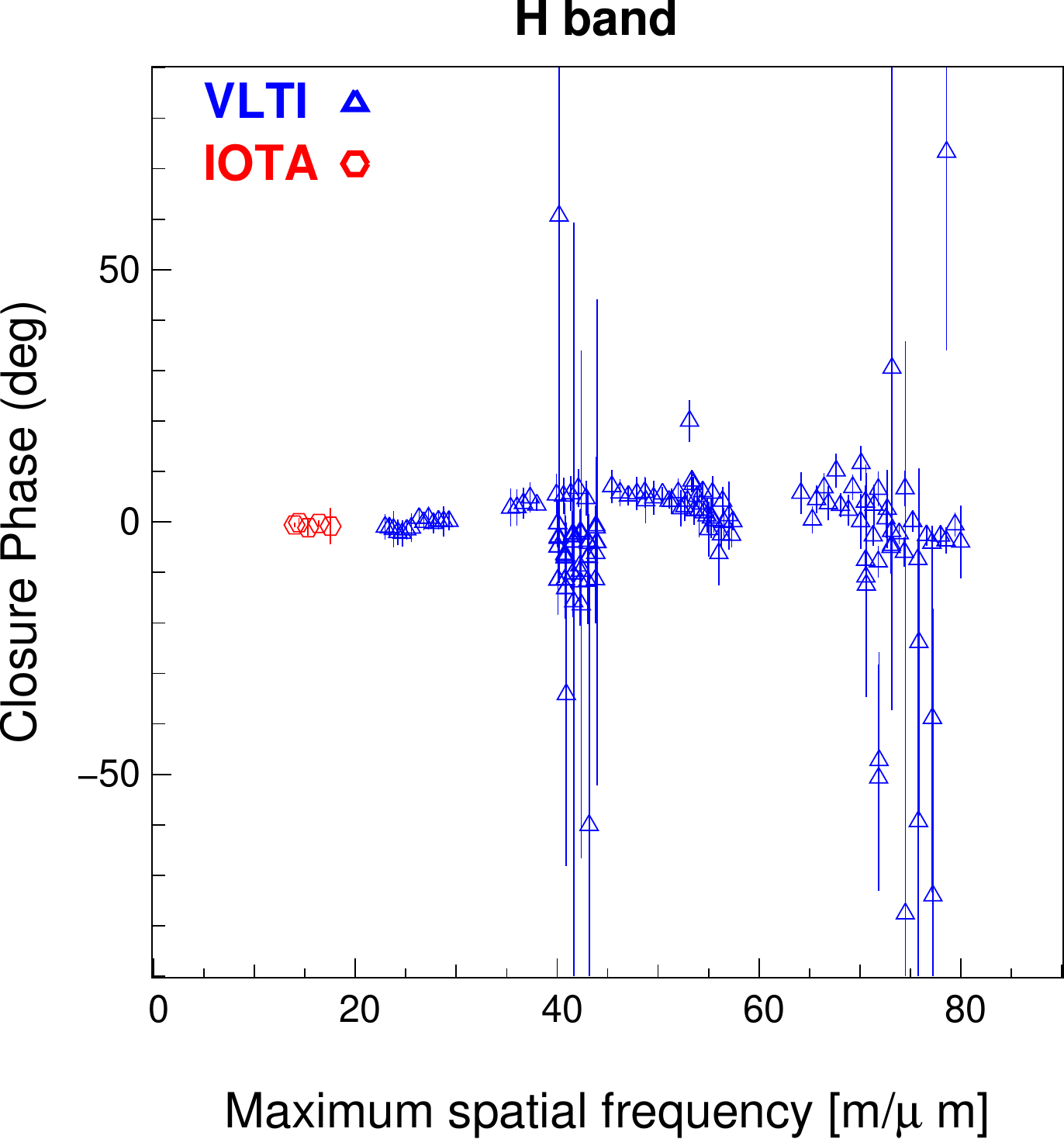}&
    \includegraphics[width=6.5cm]{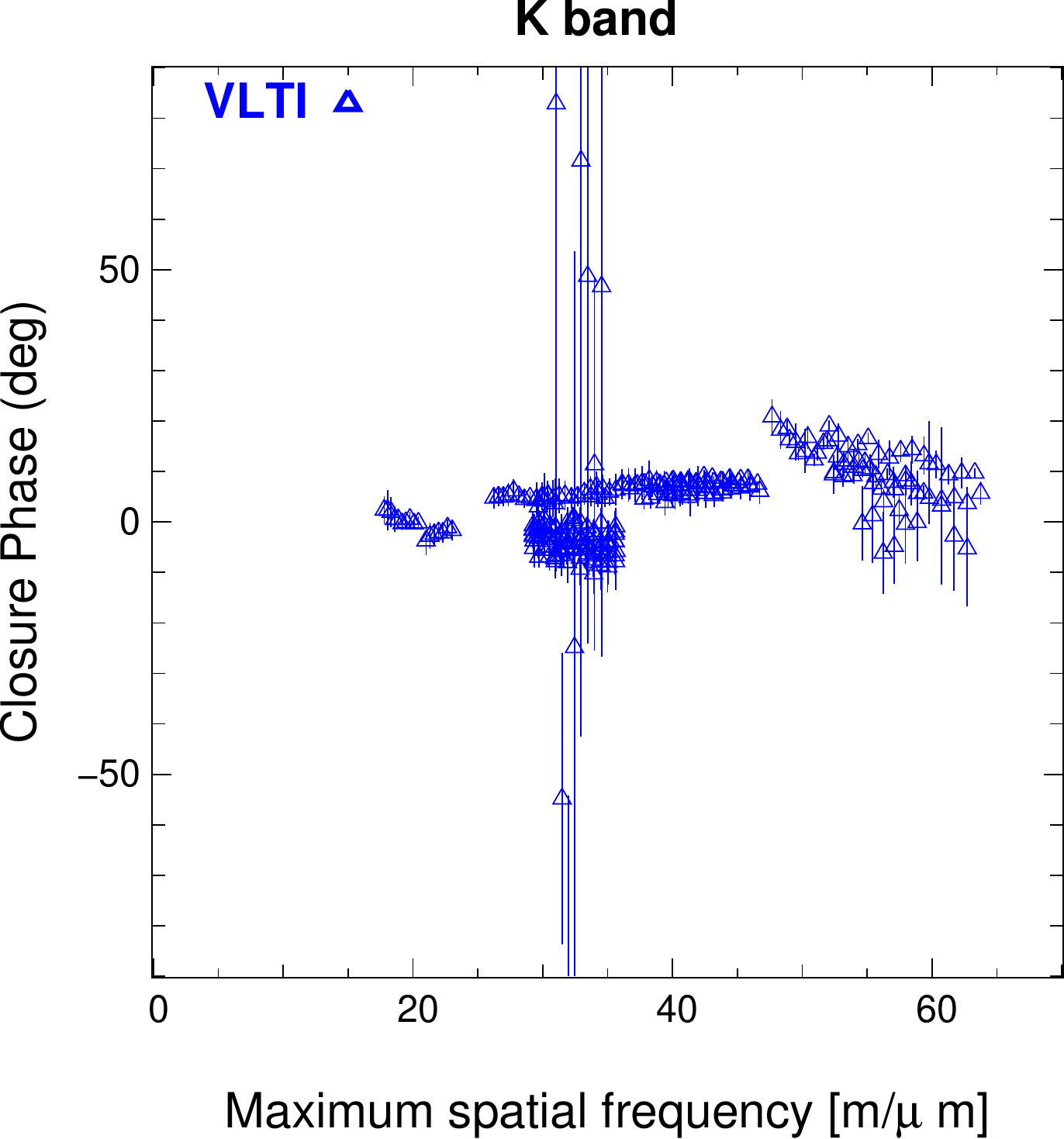}
  \end{tabular}
  \caption{Squared visibilities (up) and closure phases (bottom) in the
    $H$ (left) and $K$ (right) bands. The different interferometers are
    plotted in different colors and symbols.}
  \label{fig:uvdata}
\end{figure*}
\begin{figure*}[p]
  \centering
  \begin{tabular}{cc}
    \includegraphics[width=6.5cm]{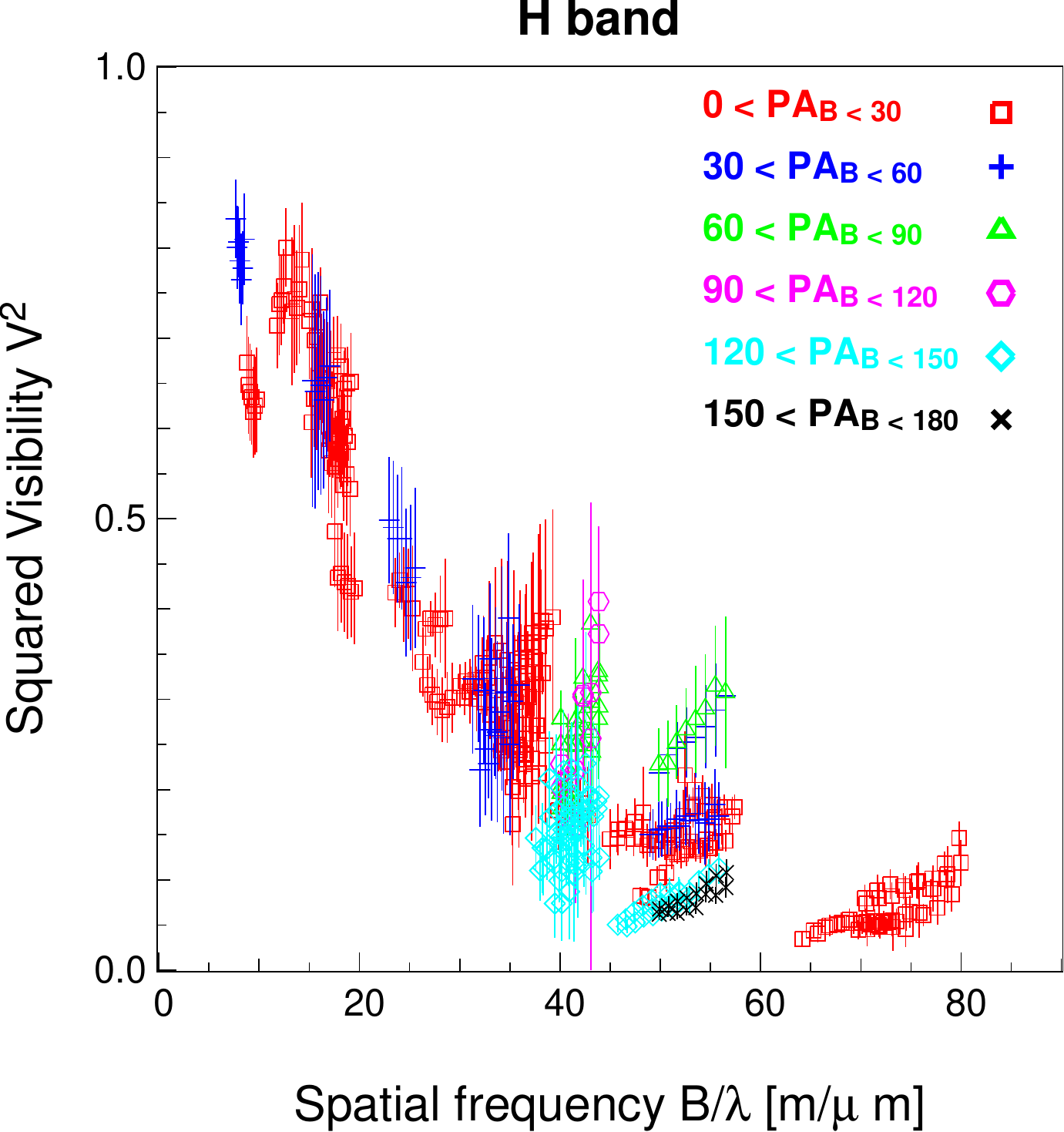}&
    \includegraphics[width=6.5cm]{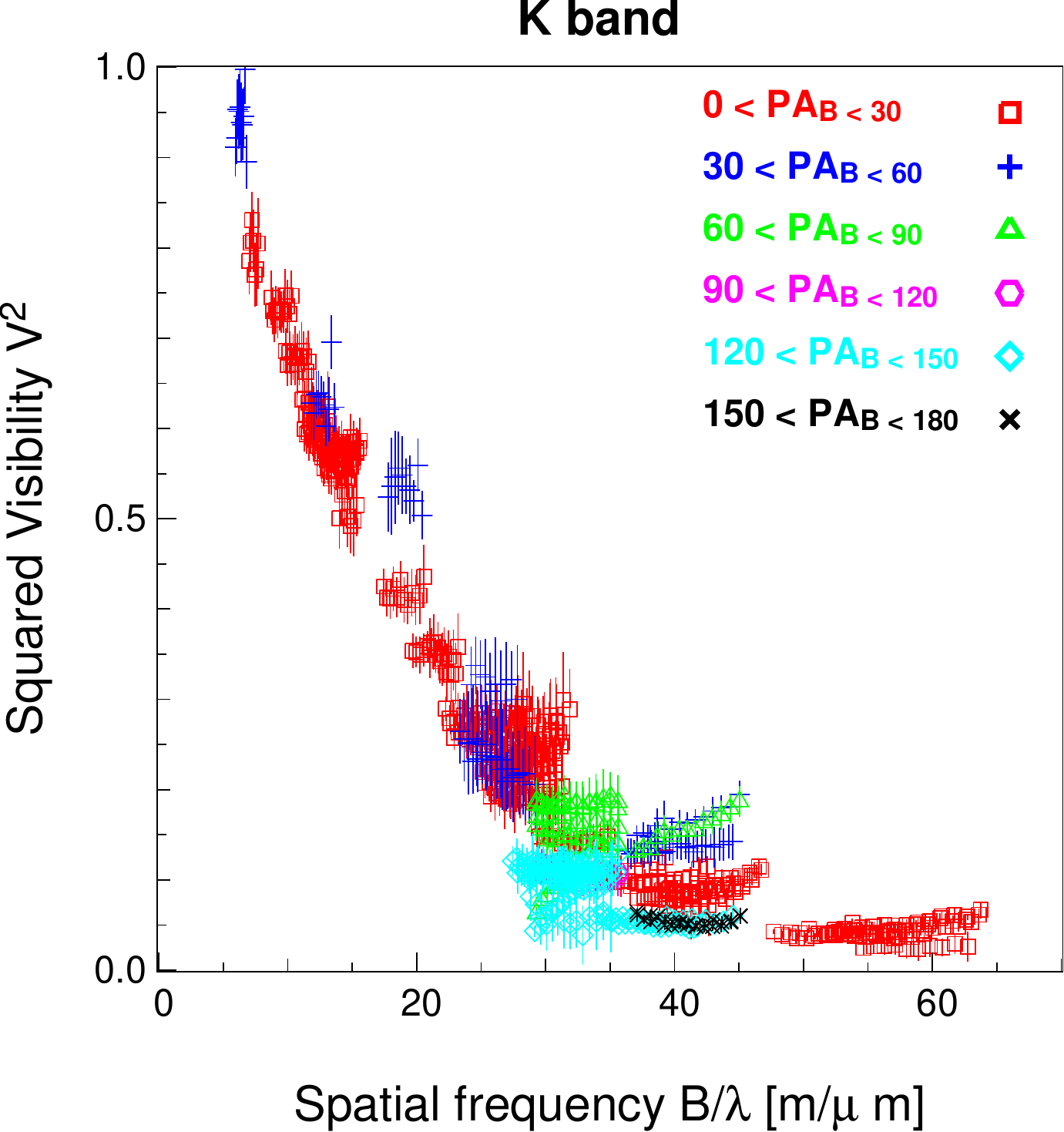}
  \end{tabular}
  \caption{Squared visibilities in the $H$ (left) and $K$ (right) bands
  for the VLTI/AMBER data. The different colors and symbols correspond to
  different baseline position angle ranges.}
  \label{fig:uvdataVLTI}
\end{figure*}
The data used in this article comes from several interferometers. The
name of these facilities, the observation date, the projected
baselines, the number of measurements, the spectral dispersion, and the
references have been summarized in Table~\ref{tab:data}. The \uv
plane in the $H$ and $K$ bands is plotted in Fig.~\ref{fig:uvcov}. The data (squared
visibilities and closure phases) is displayed in
Fig.~\ref{fig:uvdata}. For a better view, a zoom on the VLTI data is shown in Fig.~\ref{fig:uvdataVLTI}, with various colors for different ranges of baseline position angles.

\begin{table}[t]
  \caption{Log of the data used for the image reconstruction.}
  \label{tab:data}
  \centering
  \begin{tabular}{ll@{~~}l@{~~}ll@{~~}ll@{~~}ll@{~~}ll@{~~}l}
     \hline
    Interferom. &Obs.\ &$B_P$ (m) &\# meas. &Band &Disp. &Refs. \\
    \hline
    KI    &2003    &72-78  &4   &$K$ &BB &M05\\
    IOTA  &2003-04 &10-28  &9   &$H$ &BB &M06\\
    CHARA &2004-07 &86-313 &16  &$K$ &BB &T08\\
    VLTI/ &\multirow{2}{*}{2008}    &\multirow{2}{*}{13-128}
                   &947 &$K$ &\multirow{2}{*}{35} &\multirow{2}{*}{B10}\\
    \,AMBER &        &       &544 &$H$ & &\\
   \hline
\multicolumn{7}{l}{\emph{Notes}: $B_P$ is the projected baselines; Disp.\ is the
  spectral dispersion; }\\
   \multicolumn{7}{l}{Refs.\ are
     M05 \citep{Monnier2005}, M06 \citep{Monnier2006},}\\
   \multicolumn{7}{l}{
     T08 \citep{Tannirkulam2008}, B10 \citep{Benisty2010}}\\
  \end{tabular}
\end{table}
%


\section{The effect of the regularization weight}
\label{app:mu}

The systematic tests performed in RTM10 shows that a weight factor
$\mu$ can be associated with a regularization term to within an order of
magnitude. In this appendix, we illustrate the effect of the $\mu$
factor on the reconstructed images, therefore determining the influence
of $\mu$ and demonstrating that $\mu=10^2$ is the best value.

\begin{figure*}[p]
  \centering
  \begin{tabular}{c}
    \includegraphics[width=\hsize]{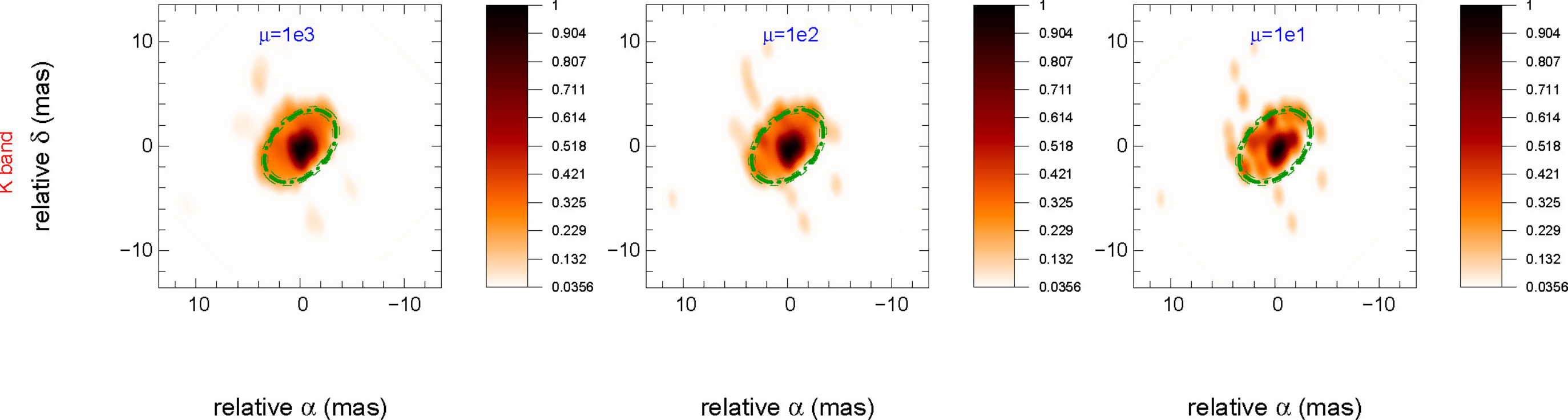} \\
    \includegraphics[width=\hsize]{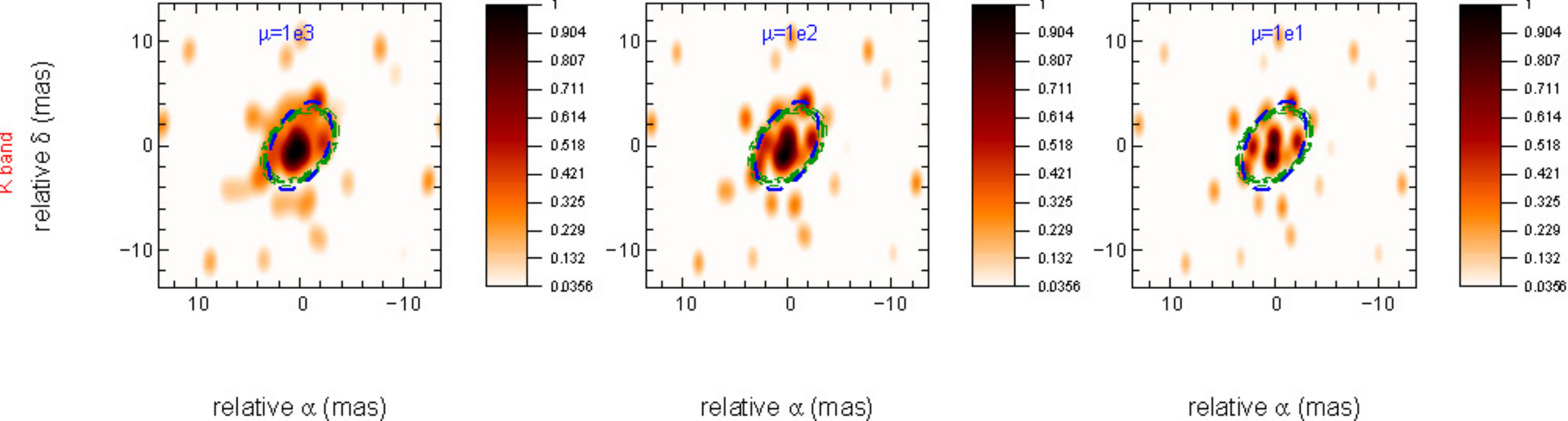}
  \end{tabular}
  \caption{Reconstructed images of \mwc\ in the $K$ band from simulated
    data of the B10 model (up) and from the real data (bottom), for 3
    different values of the weight factor $\mu$.
    conventions as in Fig.~\ref{fig:real-img-rec}.}
  \label{fig:APPAmu}
\end{figure*}

The upper part of Fig.~\ref{fig:APPAmu} presents the reconstructed
images from the B10 model in the $K$ band for 3 different values of the
weight factor $\mu$ (1000, 100, 10).  When comparing them to the model (see
the upper left corner of Fig.~\ref{fig:model-img-rec}), the
following information can be extracted:
\begin{itemize}
\item In the left part of Fig.~\ref{fig:APPAmu}, the image is
  too regularized, meaning that too much weight is put on the
  regularization term and not enough on the data. The reconstructed
  image is too smooth and several structure are not visible, because of the lack  of distinction between the external ring and the internal disk.
\item In the right part of Fig.~\ref{fig:APPAmu}, the image is insufficiently regularized, meaning that too much weight has been placed on the
  data. The reconstructed image is far more blobby than the original and the flux in
  the internal disk starts to disappear.
\item The ideal weight factor is shown in the middle part of
  Fig.~\ref{fig:APPAmu}. In this figure, all the different
  characteristics of the model are presented and it seems to provide the best
  reconstruction between the three.
\end{itemize}
The same analysis can be  performed on the reconstructed images from the
real data and are illustrated in the bottom part of
Fig.~\ref{fig:APPAmu}.

\end{document}